%% file: main.tex
\documentclass[12pt,a4paper]{article}

\usepackage[colorinlistoftodos]{todonotes}

\usepackage{ifthen} 
\newboolean{pdflatex}
\setboolean{pdflatex}{true} 

\newboolean{articletitles}
\setboolean{articletitles}{true} 

\newboolean{uprightparticles}
\setboolean{uprightparticles}{false} 

\newboolean{inbibliography}
\setboolean{inbibliography}{false} 

\def\paperauthors{LHCb collaboration} 
\def\paperasciititle{Background studies for the CODEX-b experiment: measurements and simulation} 
\def\papertitle{Background studies for the \cod experiment: measurements and simulation} 
\def\paperkeywords{{High Energy Physics}, {LHCb}} 
\def\papercopyright{\the\year\ CERN for the benefit of the LHCb collaboration} 

\def\paperlicenceurl{https://creativecommons.org/licenses/by/4.0/}

\input{preamble}
\usepackage{longtable} 
\usepackage{subfigure}

\begin{document}

\renewcommand{\thefootnote}{\fnsymbol{footnote}}
\setcounter{footnote}{1}

\input{title-LHCb-INT}

\renewcommand{\thefootnote}{\arabic{footnote}}
\setcounter{footnote}{0}



\pagestyle{plain} 
\setcounter{page}{1}
\pagenumbering{arabic}


%

\input{chapters/introduction}

\input{chapters/measurement}

\input{chapters/simulation}

\input{chapters/summary}

\input{acknowledgements}





\newpage
\addcontentsline{toc}{section}{References}
\setboolean{inbibliography}{true}
\bibliographystyle{LHCb}
\bibliography{main}

\newpage

\end{document}

%% file: preamble.tex
\usepackage[top=1in, bottom=1.25in, left=1in, right=1in]{geometry}

\usepackage{microtype}
\usepackage{lineno}  
\usepackage{xspace} 
\usepackage{caption} 

\usepackage{graphicx}  
\usepackage{subfigure}
\usepackage{rotating}
\usepackage{multirow}
\usepackage{comment}
\usepackage{color}
\usepackage{colortbl}
\usepackage{slashbox}
\usepackage{array}
\usepackage{makecell}

\graphicspath{{./figs/}} 
\DeclareGraphicsExtensions{.pdf,.PDF,png,.PNG}

\usepackage{amsmath} 
\usepackage{amssymb}
\usepackage{amsfonts}
\usepackage{upgreek} 

\newcommand*\patchAmsMathEnvironmentForLineno[1]{%
\expandafter\let\csname old#1\expandafter\endcsname\csname #1\endcsname
\expandafter\let\csname oldend#1\expandafter\endcsname\csname
end#1\endcsname
 \renewenvironment{#1}%
   {\linenomath\csname old#1\endcsname}%
   {\csname oldend#1\endcsname\endlinenomath}%
}
\newcommand*\patchBothAmsMathEnvironmentsForLineno[1]{%
  \patchAmsMathEnvironmentForLineno{#1}%
  \patchAmsMathEnvironmentForLineno{#1*}%
}
\AtBeginDocument{%
\patchBothAmsMathEnvironmentsForLineno{equation}%
\patchBothAmsMathEnvironmentsForLineno{align}%
\patchBothAmsMathEnvironmentsForLineno{flalign}%
\patchBothAmsMathEnvironmentsForLineno{alignat}%
\patchBothAmsMathEnvironmentsForLineno{gather}%
\patchBothAmsMathEnvironmentsForLineno{multline}%
\patchBothAmsMathEnvironmentsForLineno{eqnarray}%
}

\usepackage{hyperxmp}

\usepackage[pdftex,
            pdfauthor={\paperauthors},
            pdftitle={\paperasciititle},
            pdfkeywords={\paperkeywords},
           ]{hyperref}

\usepackage[all]{hypcap} 

\input{lhcb-symbols-def} 

\input{babarsym} 

\usepackage{cite} 
\usepackage{mciteplus}

%% file: lhcb-symbols-def.tex
\usepackage{xspace} 
\usepackage{upgreek}

\def\lhcb   {\mbox{LHCb}\xspace}

\def\babar  {\mbox{BaBar}\xspace}

\def\delphi {\mbox{DELPHI}\xspace}

\def\MagUp {\mbox{\em Mag\kern -0.05em Up}\xspace}

\ifthenelse{\boolean{uprightparticles}}%
{

 \def\PDelta      {\ensuremath{\Delta}\xspace}                 
 \def\PXi      {\ensuremath{\Xi}\xspace}                 
 \def\PLambda      {\ensuremath{\Lambda}\xspace}                 
 \def\PSigma      {\ensuremath{\Sigma}\xspace}                 
 \def\POmega      {\ensuremath{\Omega}\xspace}                 
 \def\PUpsilon      {\ensuremath{\Upsilon}\xspace}                 
 
 \def\PB      {\ensuremath{\mathrm{B}}\xspace}                 
                  
 \def\PD      {\ensuremath{\mathrm{D}}\xspace}

 \def\PK      {\ensuremath{\mathrm{K}}\xspace}

 \def\Pi      {\ensuremath{\mathrm{i}}\xspace}

}
{

 \mathchardef\PDelta="7101
 \mathchardef\PXi="7104
 \mathchardef\PLambda="7103
 \mathchardef\PSigma="7106
 \mathchardef\POmega="710A
 \mathchardef\PUpsilon="7107
                  
 \def\PB      {\ensuremath{B}\xspace}                 
                  
 \def\PD      {\ensuremath{D}\xspace}

 \def\PK      {\ensuremath{K}\xspace}

 \def\Pi      {\ensuremath{i}\xspace}

}

\makeatletter
\ifcase \@ptsize \relax
  \newcommand{\miniscule}{\@setfontsize\miniscule{4}{5}}
\or
  \newcommand{\miniscule}{\@setfontsize\miniscule{5}{6}}
\or
  \newcommand{\miniscule}{\@setfontsize\miniscule{5}{6}}
\fi
\makeatother

\DeclareRobustCommand{\optbar}[1]{\shortstack{{\miniscule (\rule[.5ex]{1.25em}{.18mm})}
  \\ [-.7ex] $#1$}}

\def\kaon    {{\ensuremath{\PK}}\xspace}
  \def\Kbar    {{\kern 0.2em\overline{\kern -0.2em \PK}{}}\xspace}

\def\KorKbar    {\kern 0.18em\optbar{\kern -0.18em K}{}\xspace}

\def\KL      {{\ensuremath{\kaon^0_{\mathrm{ \scriptscriptstyle L}}}}\xspace}

  \def\Dbar    {{\kern 0.2em\overline{\kern -0.2em \PD}{}}\xspace}

\def\DorDbar    {\kern 0.18em\optbar{\kern -0.18em D}{}\xspace}

\def\Bbar    {{\ensuremath{\kern 0.18em\overline{\kern -0.18em \PB}{}}}\xspace}

\def\BorBbar    {\kern 0.18em\optbar{\kern -0.18em B}{}\xspace}

  \def\Y#1S{\ensuremath{\PUpsilon{(#1S)}}\xspace}

\def\Lbar        {{\ensuremath{\kern 0.1em\overline{\kern -0.1em\PLambda}}}\xspace}
\def\LorLbar    {\kern 0.18em\optbar{\kern -0.18em \PLambda}{}\xspace}

\def\to                 {\ensuremath{\rightarrow}\xspace}

\def\AT#1     {\ensuremath{A_{\mathrm{T}}^{#1}}\xspace}           

\def\C#1      {\ensuremath{\mathcal{C}_{#1}}\xspace}                       
\def\Cp#1     {\ensuremath{\mathcal{C}_{#1}^{'}}\xspace}                    
\def\Ceff#1   {\ensuremath{\mathcal{C}_{#1}^{\mathrm{(eff)}}}\xspace}        
\def\Cpeff#1  {\ensuremath{\mathcal{C}_{#1}^{'\mathrm{(eff)}}}\xspace}       
\def\Ope#1    {\ensuremath{\mathcal{O}_{#1}}\xspace}                       
\def\Opep#1   {\ensuremath{\mathcal{O}_{#1}^{'}}\xspace}                    

\newcommand{\tev}{\ifthenelse{\boolean{inbibliography}}{\ensuremath{~T\kern -0.05em eV}}{\ensuremath{\mathrm{\,Te\kern -0.1em V}}}\xspace}
\newcommand{\gev}{\ensuremath{\mathrm{\,Ge\kern -0.1em V}}\xspace}
\newcommand{\mev}{\ensuremath{\mathrm{\,Me\kern -0.1em V}}\xspace}
\newcommand{\kev}{\ensuremath{\mathrm{\,ke\kern -0.1em V}}\xspace}
\newcommand{\ev}{\ensuremath{\mathrm{\,e\kern -0.1em V}}\xspace}
\newcommand{\gevc}{\ensuremath{{\mathrm{\,Ge\kern -0.1em V\!/}c}}\xspace}
\newcommand{\mevc}{\ensuremath{{\mathrm{\,Me\kern -0.1em V\!/}c}}\xspace}
\newcommand{\gevcc}{\ensuremath{{\mathrm{\,Ge\kern -0.1em V\!/}c^2}}\xspace}
\newcommand{\gevgevcccc}{\ensuremath{{\mathrm{\,Ge\kern -0.1em V^2\!/}c^4}}\xspace}
\newcommand{\mevcc}{\ensuremath{{\mathrm{\,Me\kern -0.1em V\!/}c^2}}\xspace}

\def\gsim{{~\raise.15em\hbox{$>$}\kern-.85em
          \lower.35em\hbox{$\sim$}~}\xspace}
\def\lsim{{~\raise.15em\hbox{$<$}\kern-.85em
          \lower.35em\hbox{$\sim$}~}\xspace}

\def\tell1  {TELL1\xspace}
\def\ukl1   {UKL1\xspace}

\newcommand{\etal}{\mbox{\itshape et al.}\xspace}

\def\cod  {CODEX-b\xspace}

%% file: babarsym.tex
\usepackage{relsize}
\def\babar{\mbox{\slshape B\kern-0.1em{\smaller A}\kern-0.1em
    B\kern-0.1em{\smaller A\kern-0.2em R}}\;}

%% file: title-LHCb-INT.tex

\begin{titlepage}

\vspace*{-1.5cm}


\vspace*{2.0cm}

{\normalfont\bfseries\boldmath\huge
\begin{center}
  \papertitle
\end{center}
}

\vspace*{0.5cm}

\begin{center}
Biplab Dey$^1$, Jongho Lee$^{2,3}$, Victor Coco$^2$ and Chang-Seong Moon$^3$ 
\bigskip\\
{\normalfont\itshape\footnotesize
$ ^1$ELTE E\"{o}tv\"{o}s Lor\'{a}nd University, Budapest, Hungary\\
$ ^2$European Organization for Nuclear Research, Geneva, Switzerland\\
$ ^3$Kyungpook National University, Daegu, South Korea
}

\vspace{0.5cm}
\today
\end{center}

\vspace*{0.5cm}

\begin{abstract}
\noindent This report presents results from a background measurement campaign for the \cod proposal undertaken in August, 2018. The data were recorded in the DELPHI side of the LHCb cavern behind a 3.2~m concrete shield wall, during Run~2 proton-proton collisions with the goal of calibrating the simulation for the full \cod detector. The maximum flux rate in the DELPHI side of the cavern was found to be around 0.6~mHz/cm$^2$ across a vertical plane just behind the shield wall, parallel to the beam line. A detailed simulation under development within the LHCb {\tt Gauss} framework is described. This includes shielding elements pertinent for \cod's acceptance -- the LHCb detector, the shield wall and cavern infrastructure. Additional flux from tracks not in the line of sight from the interaction point, but bent by the magnetic fields, are incorporated. Overall, the simulation overestimates the background flux compared to the measurement. 
Several cross-checks and avenues for further investigations are described.
\end{abstract}



\vspace{\fill}


\end{titlepage}

\pagestyle{empty}  


%

%% file: chapters/introduction.tex
\section{Introduction and motivation}
\label{sec:Introduction}

The discovery of the Higgs boson at the LHC in 2012 filled in the last missing piece of the Standard Model (SM). Apart from a few so-called ``anomalies'', mostly in the flavor sector, the SM has been a spectacularly successful framework that can account for a huge array of high precision measurements. Yet, the SM is also an incomplete theory that does not account for gravity, dark matter, observed matter-antimatter asymmetry in the universe, among other problems. New Physics (NP) searches at the LHC experiments have mostly focused on production of new particles that decay close to the collision point, and can be detected within the detector volume. However, a number of important NP scenarios predict new particles with long lifetimes. In fact, long lifetimes are generic in any theory with multiple mass scales, broken symmetries, or restricted phase-space~\cite{Curtin:2018mvb}. The SM itself contains a number of examples of such low mass, unstable but long-lived particles (LLP) such as pions, kaons, muons and neutrons. 

In keeping with these considerations, there has been a strong thrust to ensure that the High-Luminosity LHC does not miss NP signals from the LLP sector. Several new experiments, MATHUSLA~\cite{Chou:2016lxi}, FASER~\cite{Feng:2017uoz}, MilliQan~\cite{Ball:2016zrp}, SHiP~\cite{Alekhin:2015byh}, AL3X~\cite{Gligorov:2018vkc} and \cod~\cite{Gligorov:2017nwh,Aielli:2019ivi}, have been proposed within the CERN complex. A detailed discussion of the theory motivation and the complementarity between experimental approaches to detection of LLPs can be found in the recently submitted \cod Expression of Interest document~\cite{Aielli:2019ivi}.

\begin{figure}[b]
\centering
    \includegraphics[width=12cm]{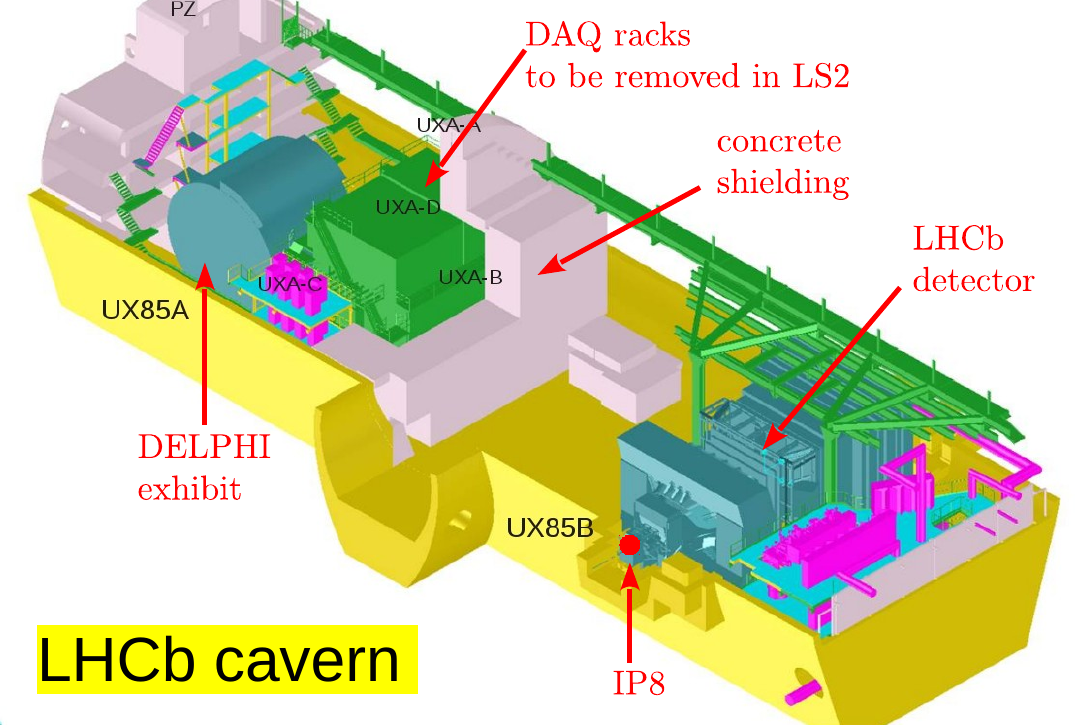}
\caption{\label{fig:cavern} 
  A schematic diagram of the LHCb cavern. The UXA-D barracks behind the concrete shield wall are shown, where \cod is proposed to be installed.  
}
\end{figure}
\subsection{A Compact Detector for Exotics (\cod) at LHCb}

Figure~\ref{fig:cavern} shows a schematic diagram of the LHCb cavern. For Run~3 operations to commence in 2022, a part of the UXA-D barracks behind the shield wall will be emptied. This includes the D1 region (ground floor) that housed the DAQ racks for Run~1-2 which will move to the newly installed Data Center~\cite{datacenter} on the surface for Run~3 onwards. The \cod proposal is to empty the entire UXA-D barracks (the D2 and D3 floors house the slow controls, HV crates, etc.), making available a volume of around $10\times10\times 10$~m$^3$ which can then be instrumented with tracking layers. The \cod volume will be stationed approximately 30~m from the collision point at Point8/LHCb. A complete description of the proposal can be found in Refs.~\cite{Gligorov:2017nwh,Aielli:2019ivi}.

%% file: chapters/measurement.tex
\section{The background measurement campaign}
\label{sec:Measurement}

\cod is envisioned as an SM background-free detector to search for long-lived particles. As shown in Fig.~\ref{fig:cavern}, the existing 3.2~m thick concrete wall separating the LHCb cavern from the DELPHI side (where \cod will be stationed) will act as the major shielding component. However, additional shielding, both passive and active, might be required to suit the needs for \cod. To understand the required shielding and perform simulation studies, data-driven calibration is required, using real collison data at IP8/LHCb. For this purpose, a campaign to measure the background flux at various locations in the DELPHI cavern (behind the shield wall) was undertaken during Run~2 operations (proton-proton collisions) in August 2018. In the first part of this document, a detailed report of this campaign is presented.

\begin{figure}
\centering
\subfigure[]{
\centering
   \includegraphics[width=0.47\columnwidth]{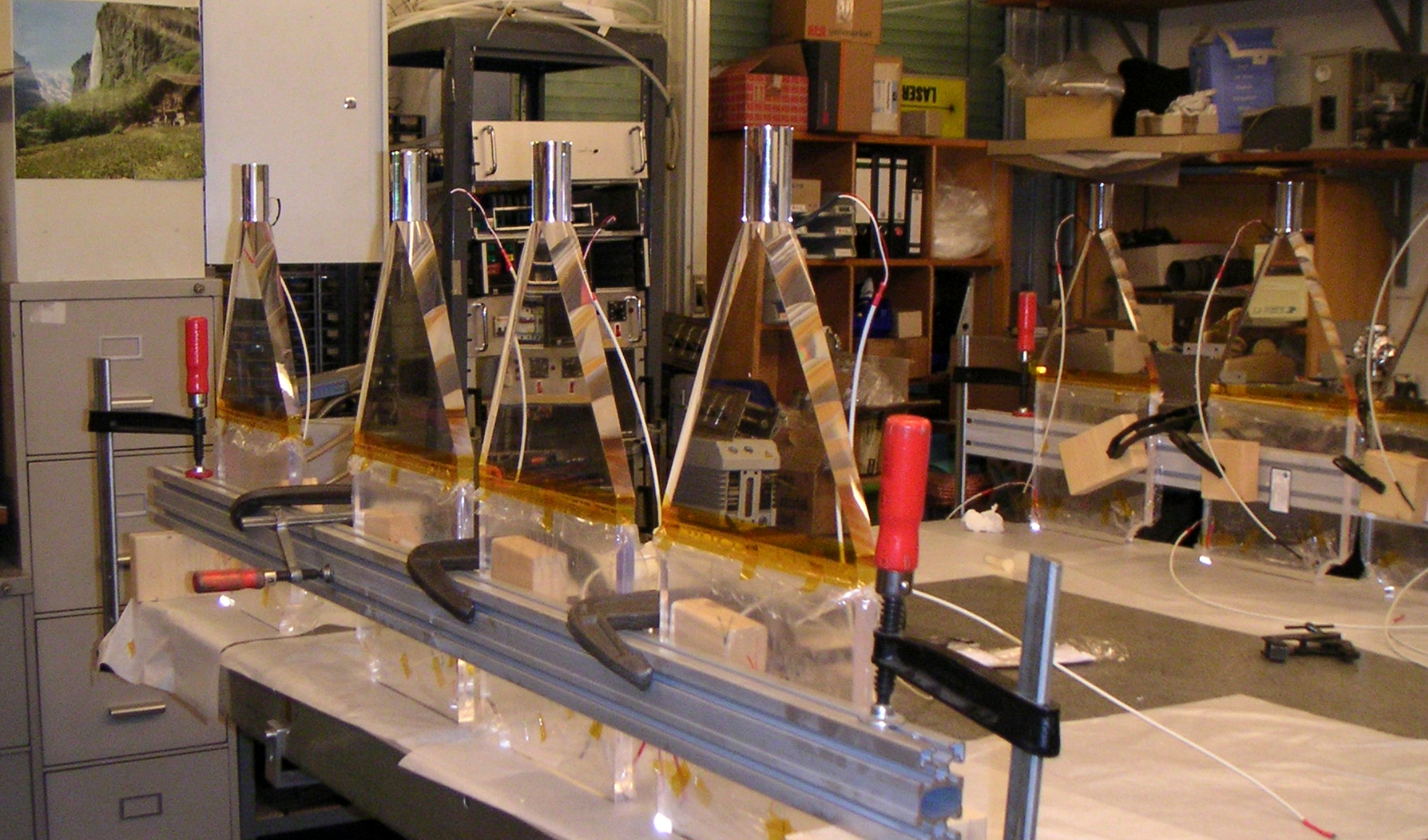}
}
\subfigure[]{
\centering
   \includegraphics[width=0.47\columnwidth]{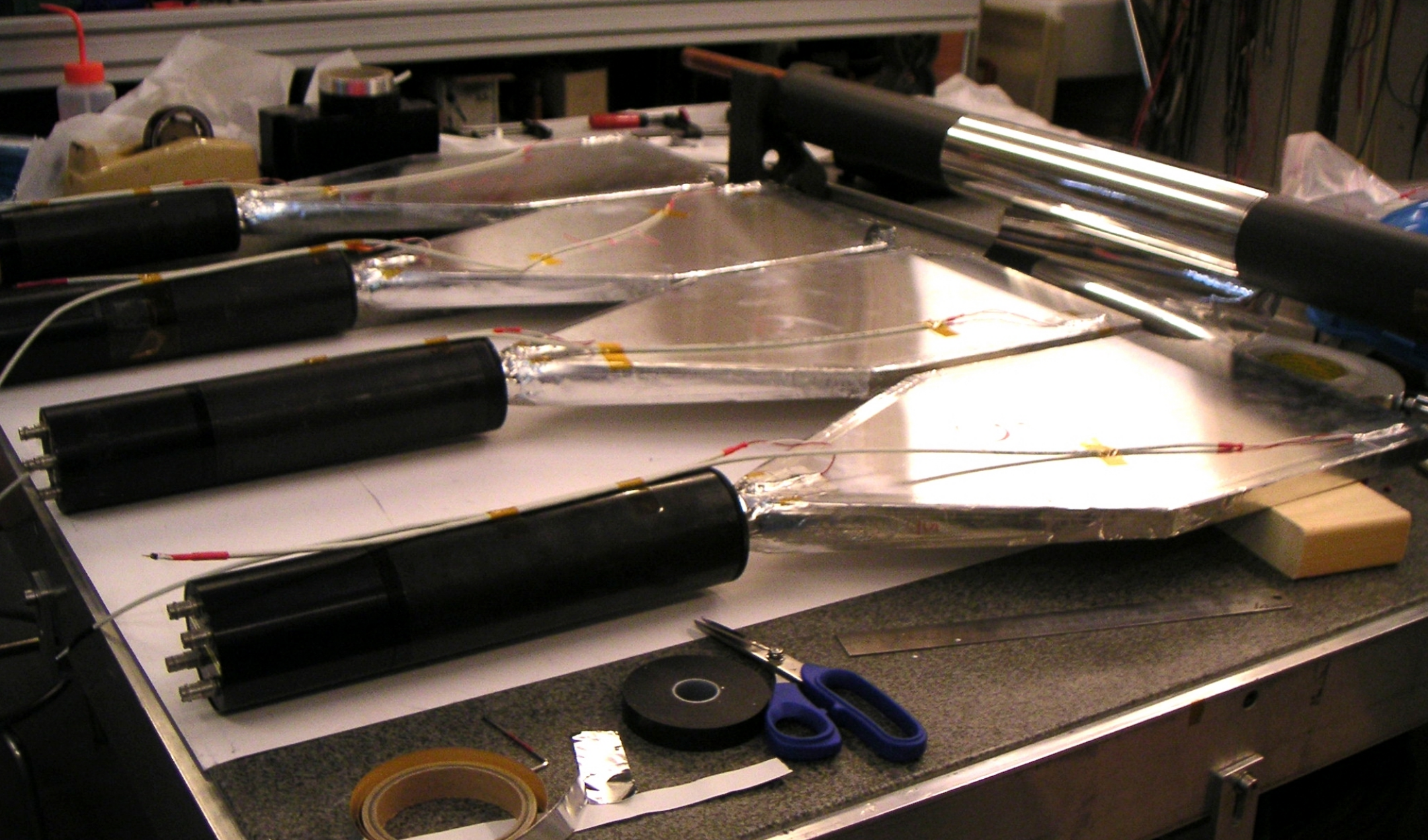} 
}
\caption{\label{fig:scint}
HeRSCheL scintillators: (a) scintillator and light-guide assembly, (b) wrapped scintillators coupled to the PMT in the casing with additional electronics.
}
\end{figure}

\subsection{The scintillators, PMTs and cosmic test-bench}
The measurement setup re-used scintillators, light-guides and photomultiplier tubes (PMT) from the HeRSCheL detector~\cite{LHCb-DP-2016-003} in LHCb. The plastic scintillating material was EJ-200~\footnote{Eljen Technology, Sweetwater, Texas, United States (\href{http://www.eljentechnology.com}{\tt http://www.eljentechnology.com}).} of volume $300\times300\times20$~mm$^3$, and the light-guides providing the coupling to the PMTs were made of Plexiglass. The scintillators and light-guides were wrapped in light-protecting aluminium foil. Each light-guide was coupled to a Hamamatsu R1828-01 $2^{\prime\prime}$ PMT chosen due to its large range of gain adjustment and relatively fast rise time of $1.3$~ns compared to the coincidence trigger time window of 5~ns (see Sec.~\ref{sec:trigger}) and nominally to the LHC collision bunch separation interval of 25~ns. Figure~\ref{fig:scint} shows the scintillator-PMT assembly~\cite{LHCb-DP-2016-003}.

\begin{figure}
\centering
\subfigure[]{
\centering
   \includegraphics[width=0.47\columnwidth]{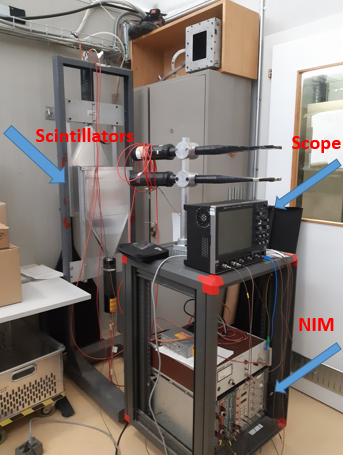}
}
\subfigure[]{
\centering
   \includegraphics[width=0.34\columnwidth]{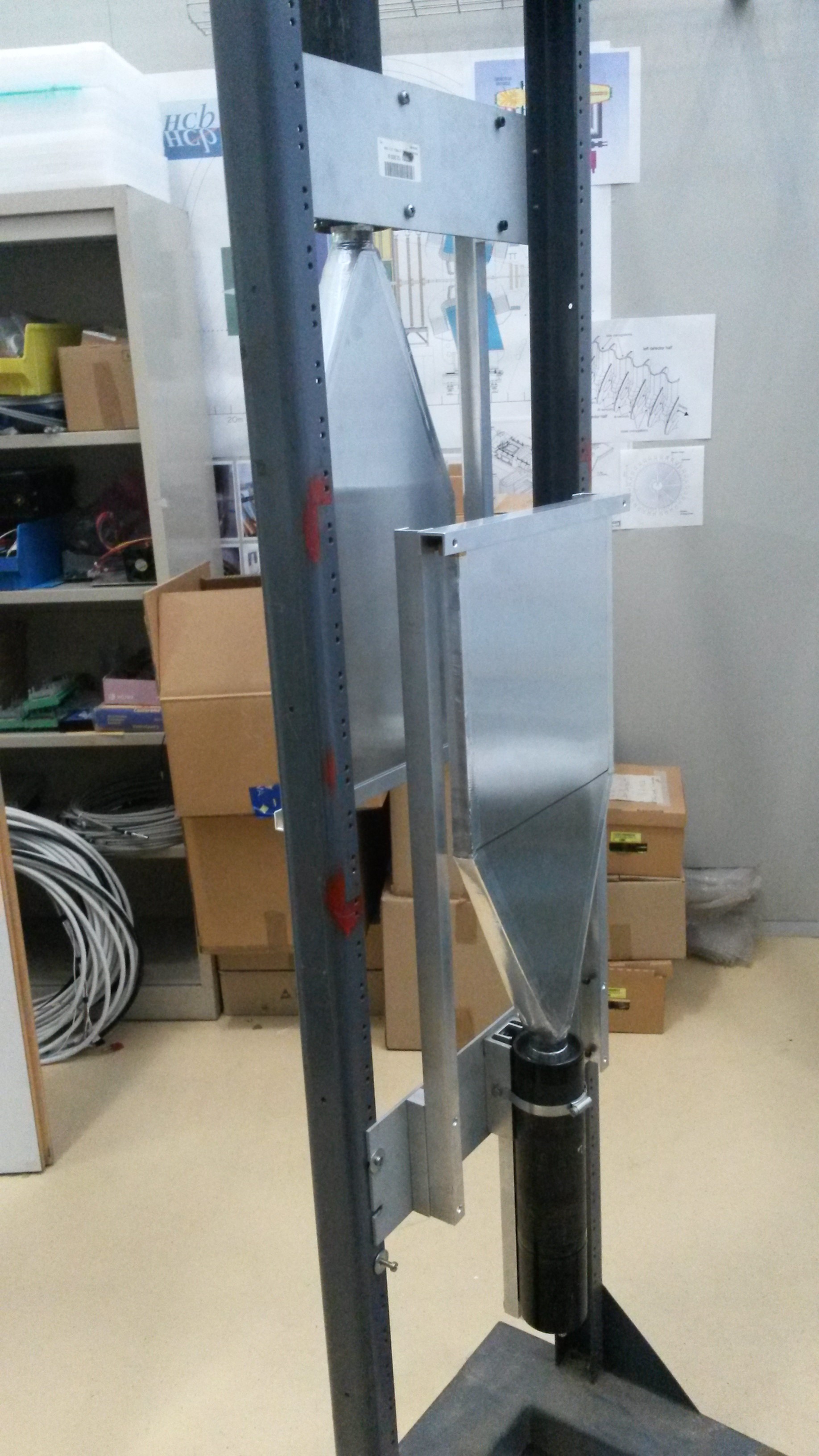} 
}
\caption{\label{fig:test_bench} Test-bench assembly in the VeloPix laboratory showing the (a) HeRSCheL scintillators, the DAQ system comprising a NIM crate and oscilloscope. The cosmic stand used for initial tests is also shown including the trigger scintillators in black, (b) close-up look at the mechanical stand.
}
\end{figure}

\begin{figure}
\centering
\subfigure[]{
\centering
   \includegraphics[width=0.295\columnwidth]{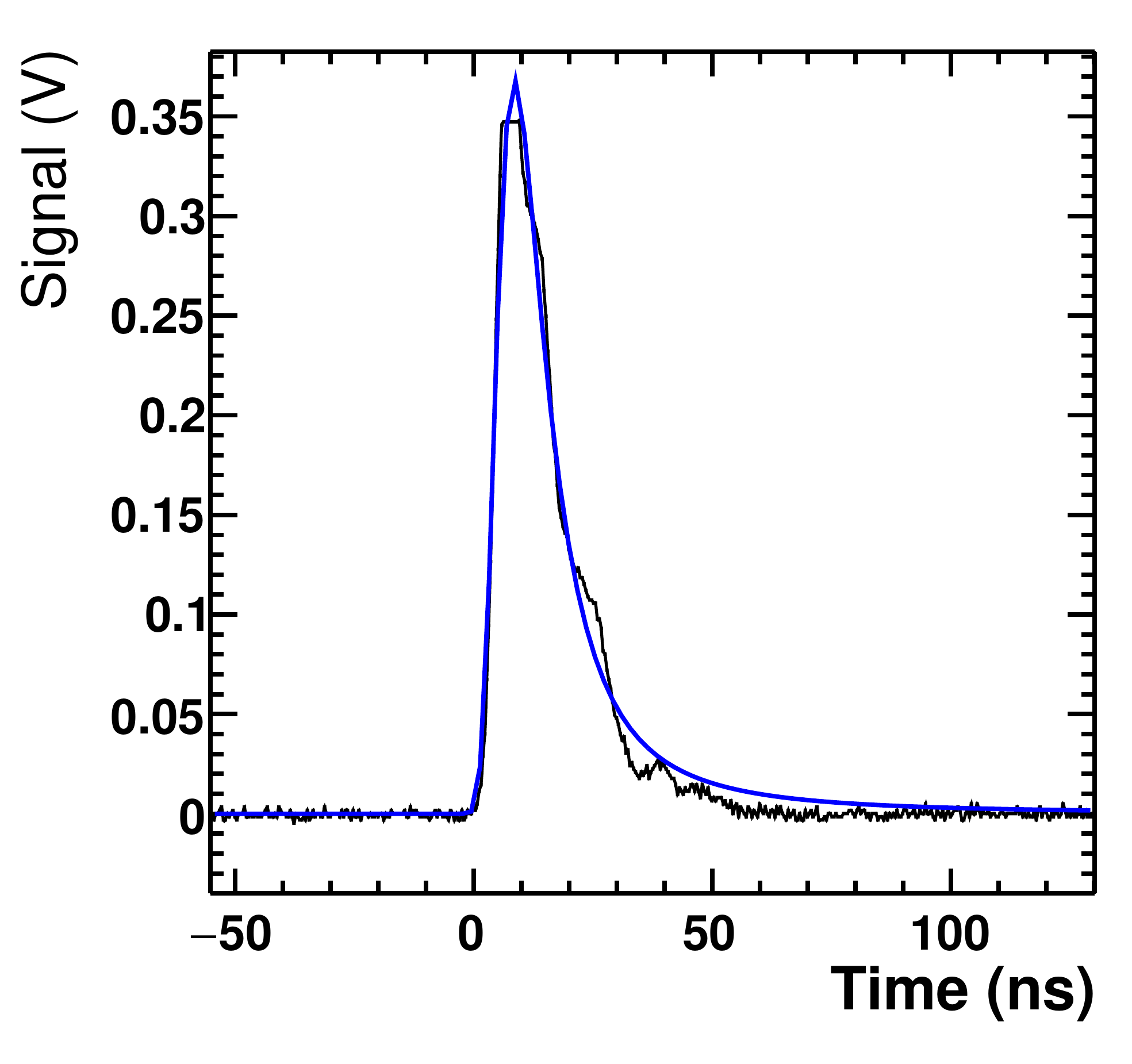} 
}
\subfigure[]{
\centering
   \includegraphics[width=0.295\columnwidth]{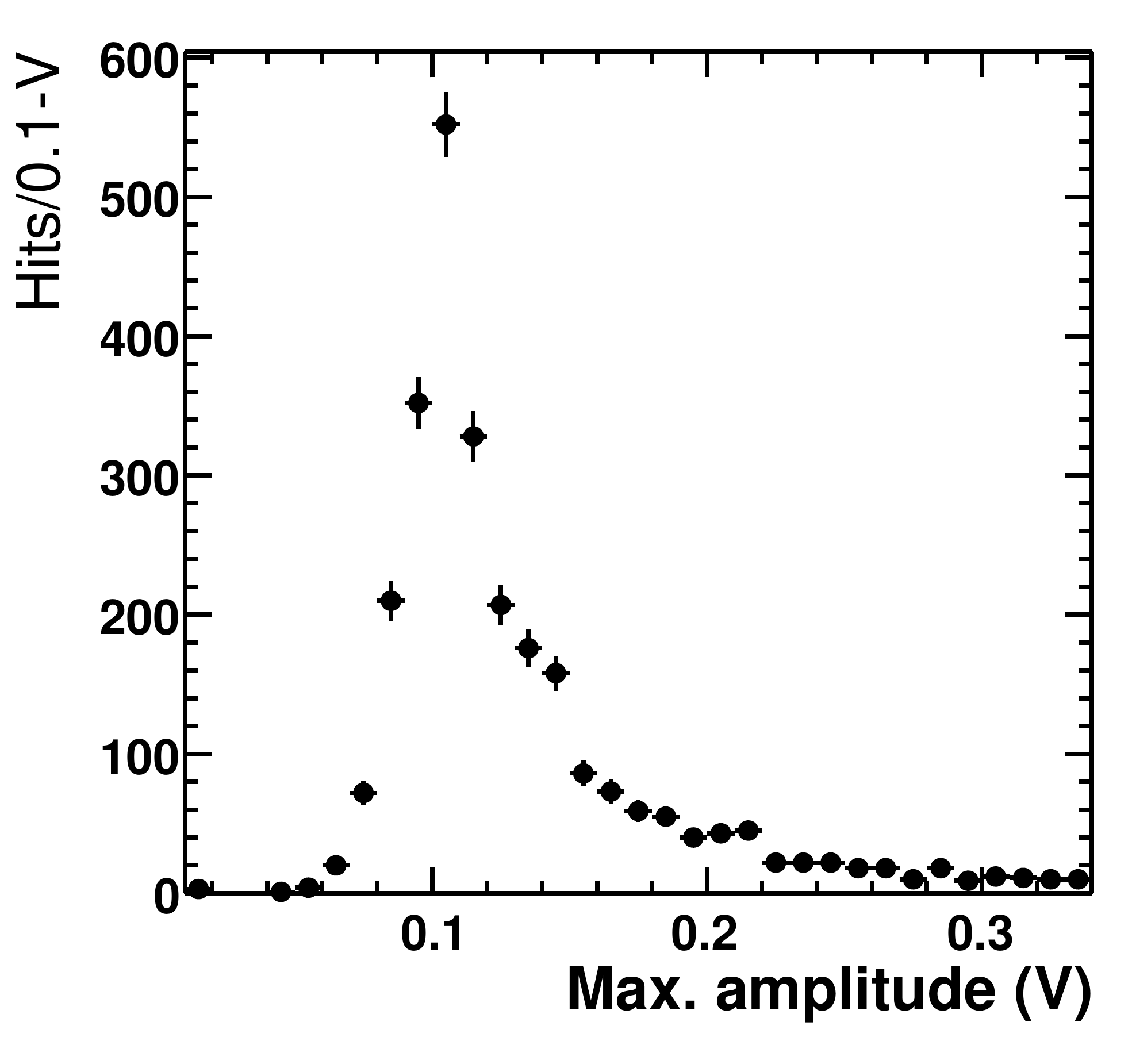}
}
\subfigure[]{
\centering
   \includegraphics[width=0.295\columnwidth]{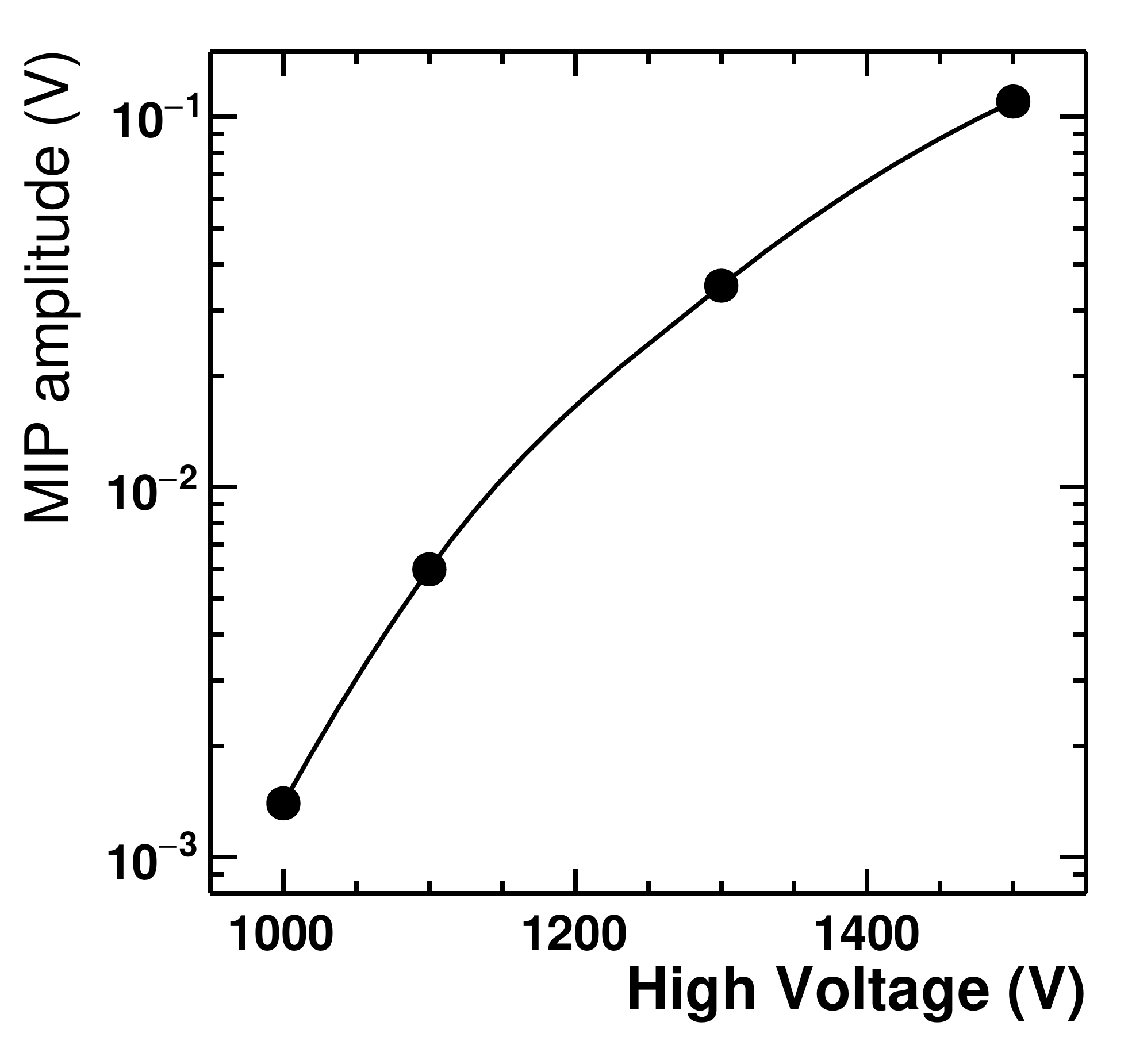}
}
\caption{\label{fig:test_bench_calib} Calibration using the cosmic test-bench assembly: (a) single event signal seen on oscilloscope for HV=1.5~kV setting along with a fit to a Landau profile shown in blue, (b) distribution of the maximum amplitudes showing the MIP amplitude to be $\approx 110$~mV for HV=1.5~kV, and (c) variation of the MIP amplitude with HV setting. All voltages are negated here for ease of representation.
}
\end{figure}

Figure~\ref{fig:test_bench} shows the test-bench assembly in the VeloPix laboratory area on the surface in Meyrin, CERN. It includes a vertical iron mechanical stand holding the wrapped scintillator pair and the NIM crate power supply providing -1.5kV HV, with an additional -350V bias voltage. The horizontal distance between the two scintillators was around 11.5~cm. The DAQ system utilized an oscilloscope (LeCroy WaveRunner) with extended functions (autosave waveforms, coincidence logic, \etal). Before transporting the setup to the underground at Point8, it was tested and calibrated with a cosmic stand shown in Fig.~\ref{fig:test_bench}a, where the black horizontal trigger scintillators provide triggers for the cosmics. Figure~\ref{fig:test_bench_calib} shows the calibration procedure. The oscilloscope signal for a single event for HV set at -1.5~kV is shown in Fig.~\ref{fig:test_bench_calib}a, along with a fit to a Landau profile (blue curve). The distribution of the maximum amplitudes for this HV setting is shown in Fig.~\ref{fig:test_bench_calib}b, where the MIP (minimum ionising particle) signal is seen at around $-110$~mV. The variation of the MIP signal with the HV setting is shown in Fig.~\ref{fig:test_bench_calib}c. For the measurements, the HV=-1.5~kV setting was chosen, with MIP discrimination threshold set as -30~mV on the oscilloscope. The MIP hit efficiency for the scintillators with was checked to be $>95\%$ from the cosmics setup.

\subsection{Trigger}
\label{sec:trigger}
\begin{figure}
\centering
    \includegraphics[width=16cm]{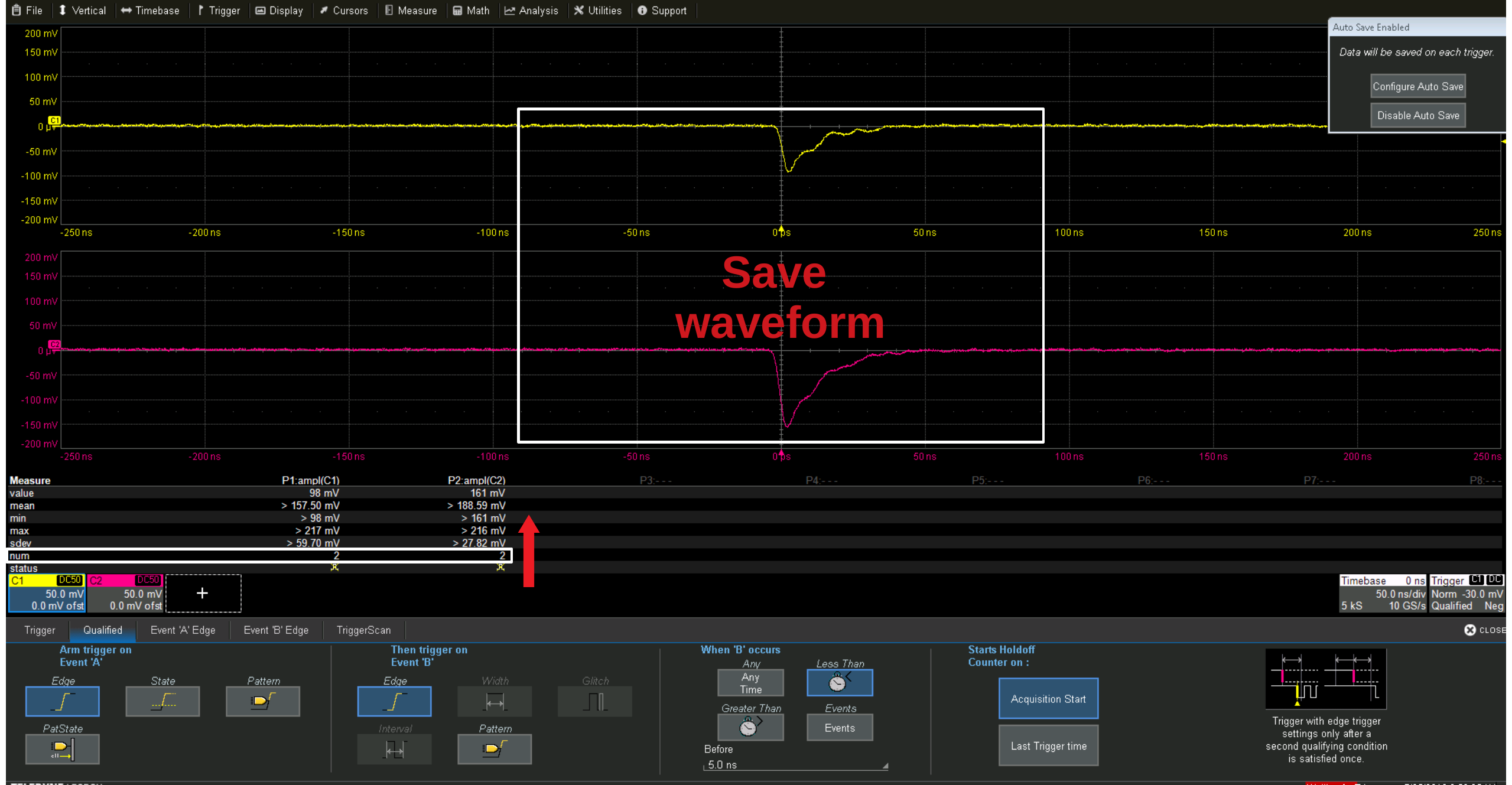}
\caption{\label{fig:waveform}
    The trigger setup using coincidence occurence of signals from the two scintillator PMTs within 5~ns. 
}
\end{figure}

A simple 2-fold coincidence between the two scintillators was used as a trigger. The time-window for the coincidence was set to be 5~ns, considerably lower than the 25~ns collision interval, so that spillover effects can be neglected. The simulation described in Sec.~\ref{sec:Simulation} also confirmed that no collision event produced more than a single hit on the scintillator planes, so that occurrence of multiple hits within the trigger window being counted as one, was negligible. The trigger was not synchronized with the collisions in any fashion. The oscilloscope automatically saved two waveforms from each scintillator, as shown in Fig.~\ref{fig:waveform}, along with a timestamp for every MIP hit event (not to be confused with collision events). This timestamp is important to correlate with the beam status during data-taking. The beam is typically on for hours and synchronising the trigger with the bunch crossings is not the goal here, but a rough comparison as to when the beam is on or off.

\subsection{Measurement positions and configurations on the D3 platform}

\begin{figure}
\centering
    \includegraphics[width=16cm]{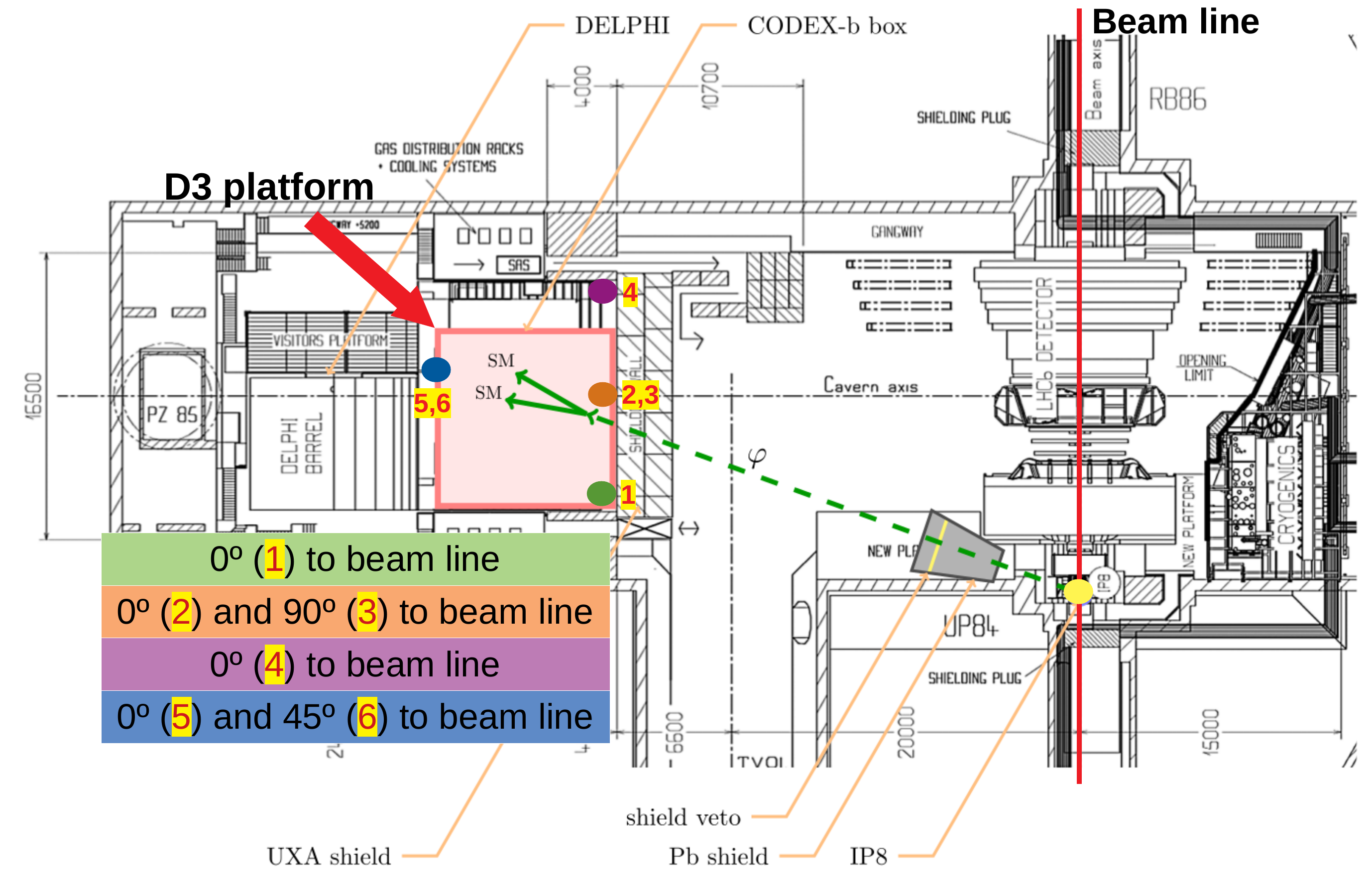}
\caption{\label{fig:posconfig}
    The four measurement positions on the D3 level inside the LHCb cavern. The configurations are labelled from P1-P6.
}
\end{figure}

\begin{figure}
\centering
\subfigure[]{
\centering
   \includegraphics[width=0.4\columnwidth]{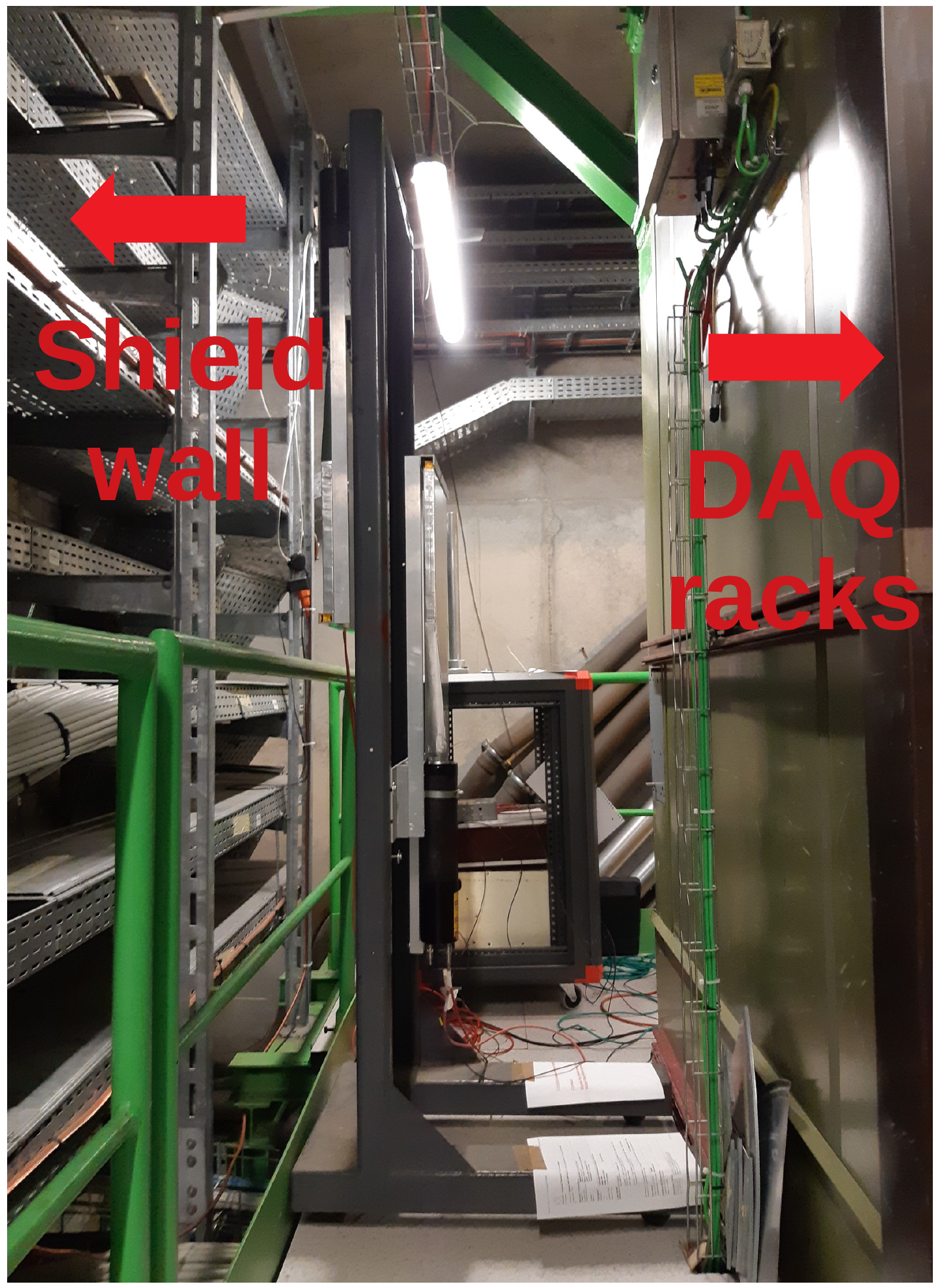} 
}
\subfigure[]{
\centering
   \includegraphics[width=0.4\columnwidth]{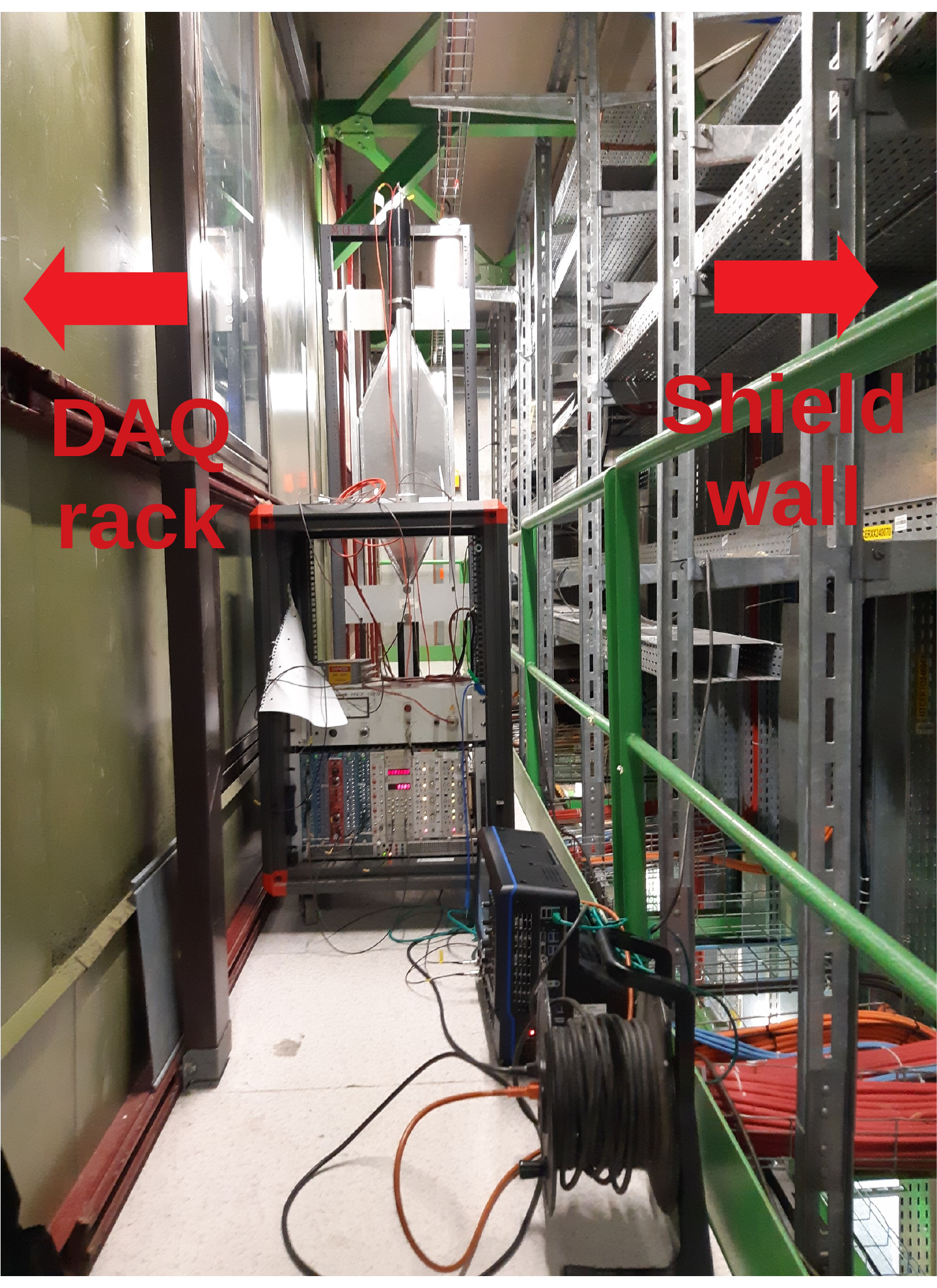}
}
\subfigure[]{
\centering
   \includegraphics[width=0.4\columnwidth]{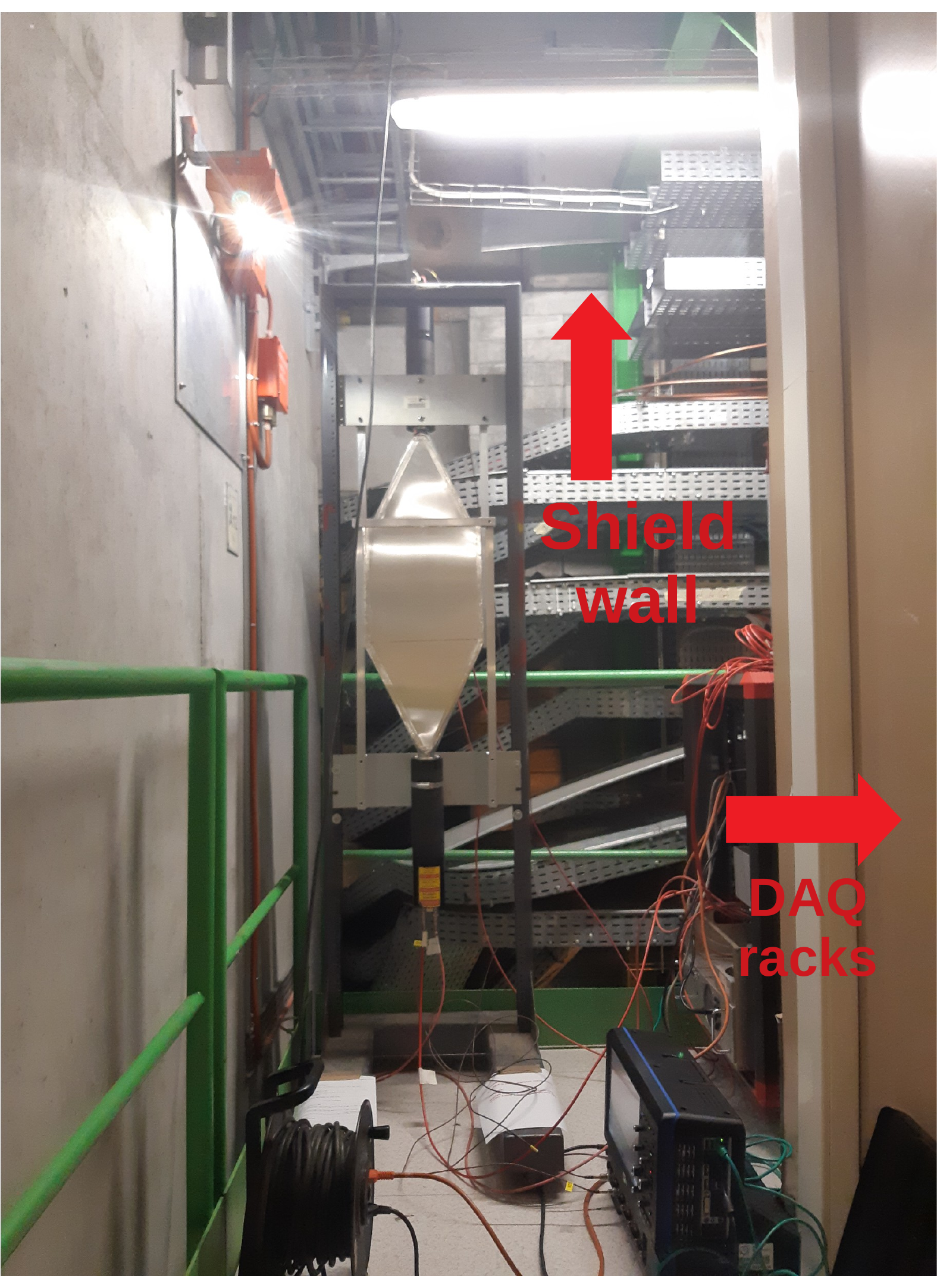} 
}
\subfigure[]{
\centering
   \includegraphics[width=0.4\columnwidth]{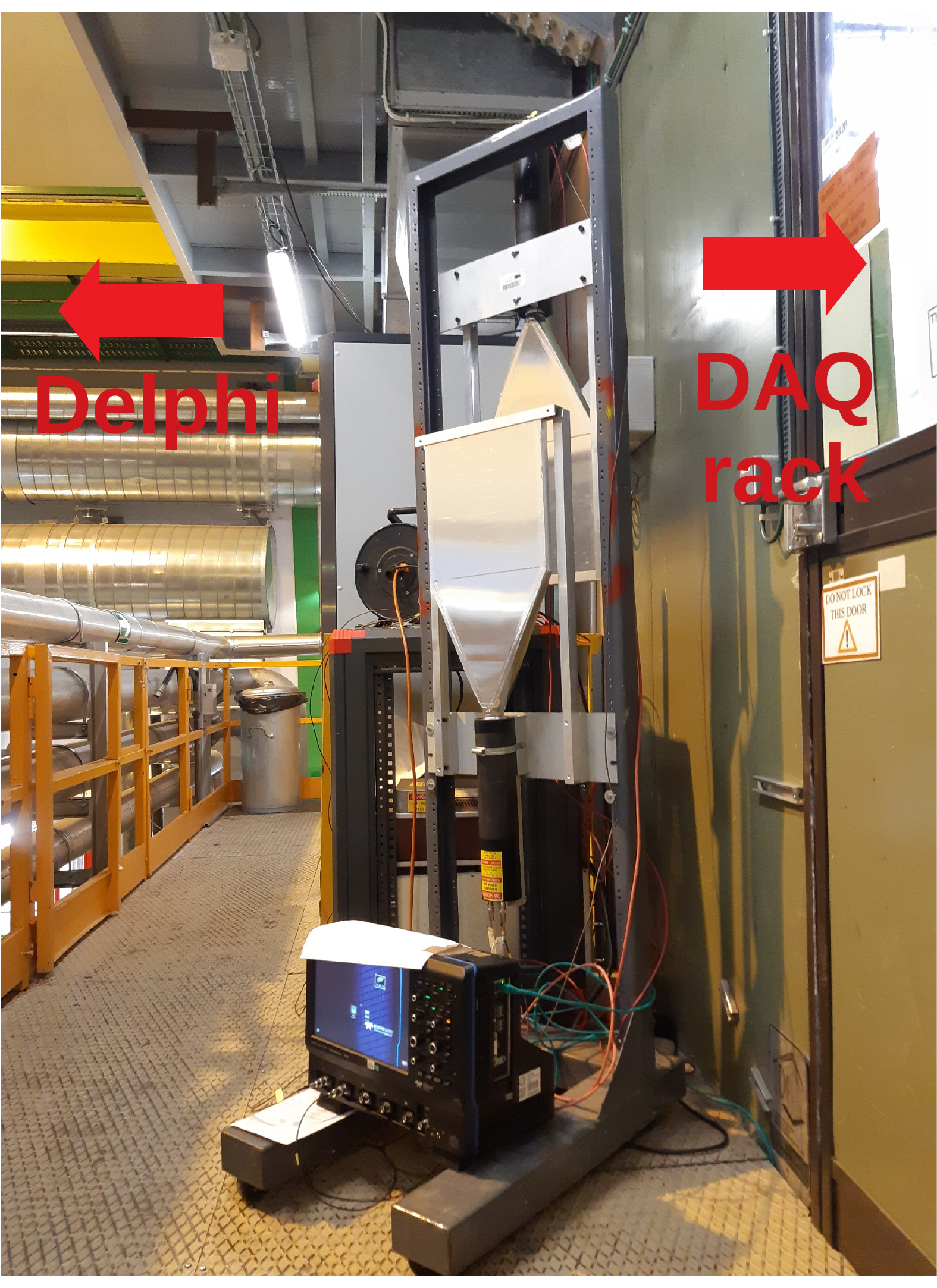} 
}
\caption{\label{fig:det_equip}
    Photos of the equipment setup at various positions on the D3 platform: (a) P1, (b) P3, (c) P4, and (d) P6.
}
\end{figure}

The background measurements were performed in the LHCb cavern on D3 platform level, just behind the concrete shield wall, on the access side. The equipment was set up at 3 positions on the passerelle between the DAQ racks and the concrete shield wall, one position between the \delphi exhibit and DAQ racks. For orientation, the scintillator stand was mostly parallel to the beam line but was also rotated $45^{\circ}$ and perpendicular to the beam line. Figure~\ref{fig:posconfig} shows the positions and configurations for the measurements, and Fig.~\ref{fig:det_equip} shows pictures of the equipment at these positions on the D3 platform.

\subsection{Results}

\begin{figure}
\centering
    \includegraphics[width=16cm]{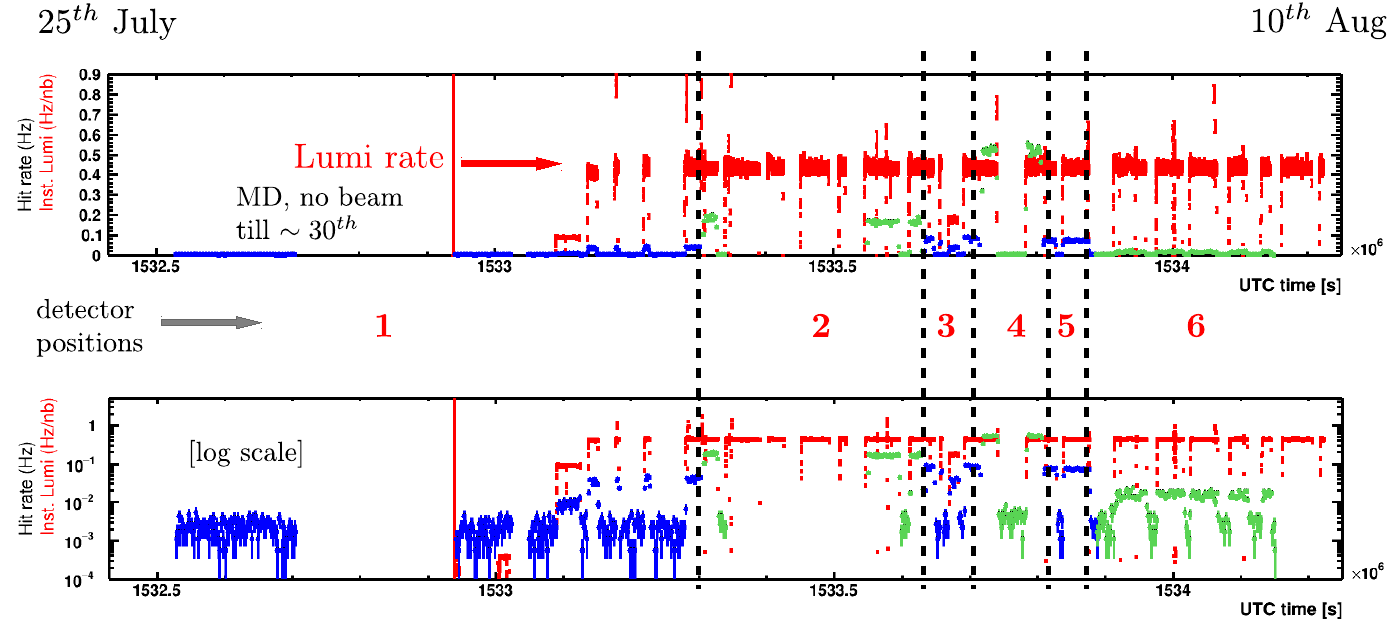}
\caption{\label{fig:results} The hit rate for the measurements based on six detector positions/configurations linear (top) and logarithmic (bottom) scale. The red dots indicate the luminosity at \lhcb, while the alternating blue and green dots indicate the hit rates at the six positions/configurations (also marked).}
\end{figure}

The measurement campaign spanned over 17 days between $25^{th}$ July and $10^{th}$ August, 2018. There were 52,036 recorded triggers during the run. The LHCb instantaneous luminosity rate was stable during the measurement. Figure~\ref{fig:results} shows the main results from the measurement campaign. The red dots correspond to the instantaneous luminosity obtained from the LHCb online database~\footnote{\href{https://lbrundb.cern.ch/}{\tt https://lbrundb.cern.ch/}} in Hz/nb. The green and blue dots show the hit rate in Hz, alternating between the six different configurations/positions, for better visibility. The plots are shown in both normal and logarithmic scales. There was no beam till July 30th because of machine development (MD) and an inadvertent power loss occurred during this initial phase.

\begin{table}
\begin{center}
\begin{tabular}{c|l|r}
  Position & \hspace{2cm}Description & Hit rate [mHz] \\
  \hline \hline
   P1 & shield, right corner, $\parallel$ to beam& $1.99\pm0.07$ \\ \hline
   P2 & shield, center, $\parallel$ to beam&  $2.76\pm 0.03$ \\ \hline
   P3 & shield, center, $\perp$ to beam& $ 2.26\pm 0.03$ \\ \hline
   P4 & shield, left corner, $\parallel$ to beam& $ 3.11\pm 0.03$ \\ \hline
   P5 & shield + D3 racks, center, $\parallel$ to beam& $ 1.95\pm 0.03$ \\ \hline
   P6 & shield + D3 racks, center, $45^\circ$ to beam& $ 2.22\pm $ 0.02\\ \hline
\end{tabular}
\caption{\label{table:rate_no_beam}
    Background hit rates based on each configuration when the beam is off.
}
\end{center}
\end{table}

Table~\ref{table:rate_no_beam} lists the hit rate from ambient background in periods without beam. The average hit rate at each position and configuration is 2~mHz.  This indicates that the ambient background can be considered negligible for this measurement. Table~\ref{table:rate_stable_beam} lists the hit rate during stable beam. The rate is non-negligible, even for a small area of $300\times300$~mm$^2$. The rate increases from P1$\to$P2$\to$P4, which, from Fig.~\ref{fig:posconfig} indicates that the downstream region sees more activity. This dependence on the $\eta$ will later shown to be reproducible in the simulation. Further, comparing the rate at P2 with P5, behind the DAQ racks, the racks are seen to add shield material. Finally, comparing the rate at P5 and P6, for the angular scan, the flux depends on the orientation with respect to the beam direction, as expected. 

From the measurement at P4, the maximal flux rate in the DELPHI side of the cavern, was found to be around 0.57~mHz/cm$^2$ across a vertical plane just behind the shield wall, parallel to the beam line.

\begin{table}
\begin{center}
\begin{tabular}{c|l|r}
  Position & \hspace{0.9cm}Description & Hit rate [mHz] \\
  \hline \hline
   P1 & shield, right corner, $\parallel$ to beam & $ 38.99 \pm 0.99 $\\ \hline
   P2 & shield, center, $\parallel$ to beam& $ 167.10 \pm 1.43$ \\ \hline
   P3 & shield, center, $\perp$ to beam& $ 82.81 \pm 1.55 $ \\ \hline
   P4 & shield, left corner, $\parallel$ to beam& $ 517.45 \pm 3.52 $ \\ \hline
   P5 & shield + D3 racks, center, $\parallel$ to beam& $ 73.58 \pm 1.18 $ \\ \hline
   P6 & shield + D3 racks, center, $45^\circ$ to beam& $ 15.71 \pm 0.33 $ \\ \hline
\end{tabular}
\caption{\label{table:rate_stable_beam}
    Average hit rates measured during stable beam, at the six positions/configurations.
}
\end{center}
\end{table}

%% file: chapters/simulation.tex
\section{Simulation}
\label{sec:Simulation}

An important aspect of the \cod proposal is the prospect of tagging events in \cod with LHCb activity, pertinent for exotic Higgs decays, for example (see Sec.~VD in Ref.~\cite{Aielli:2019ivi}). Therefore it is convenient to essentially treat \cod as a sub-detector for LHCb, the distance from IP8 to \cod being of the same order as to the muon stations at the downstream end of the LHCb detector. To understand the flux that enters the DELPHI cavern, the following features are critical
\begin{enumerate}
\item the 3.2~m thick concrete shield wall;
\item the concrete cavern wall immediately surrounding the IP that lies in the \cod acceptance;
\item effects of the various magnetic fields from the main LHCb magnet as well as the collimator magnets;
\item any other material that lies in the \cod acceptance, that can provide shielding;
\item backgrounds from non-IP8-collision sources due to the LHC machine (Machine Induced Background). 
\end{enumerate}
Items 1-3 are fully accounted for in this report, while items 4 and 5 are discussed in Secs.~\ref{sec:additional_mat} and \ref{sec:mib}, respectively.

\begin{figure}
\centering
\subfigure[]{
\centering
    \includegraphics[width=0.5\textwidth]{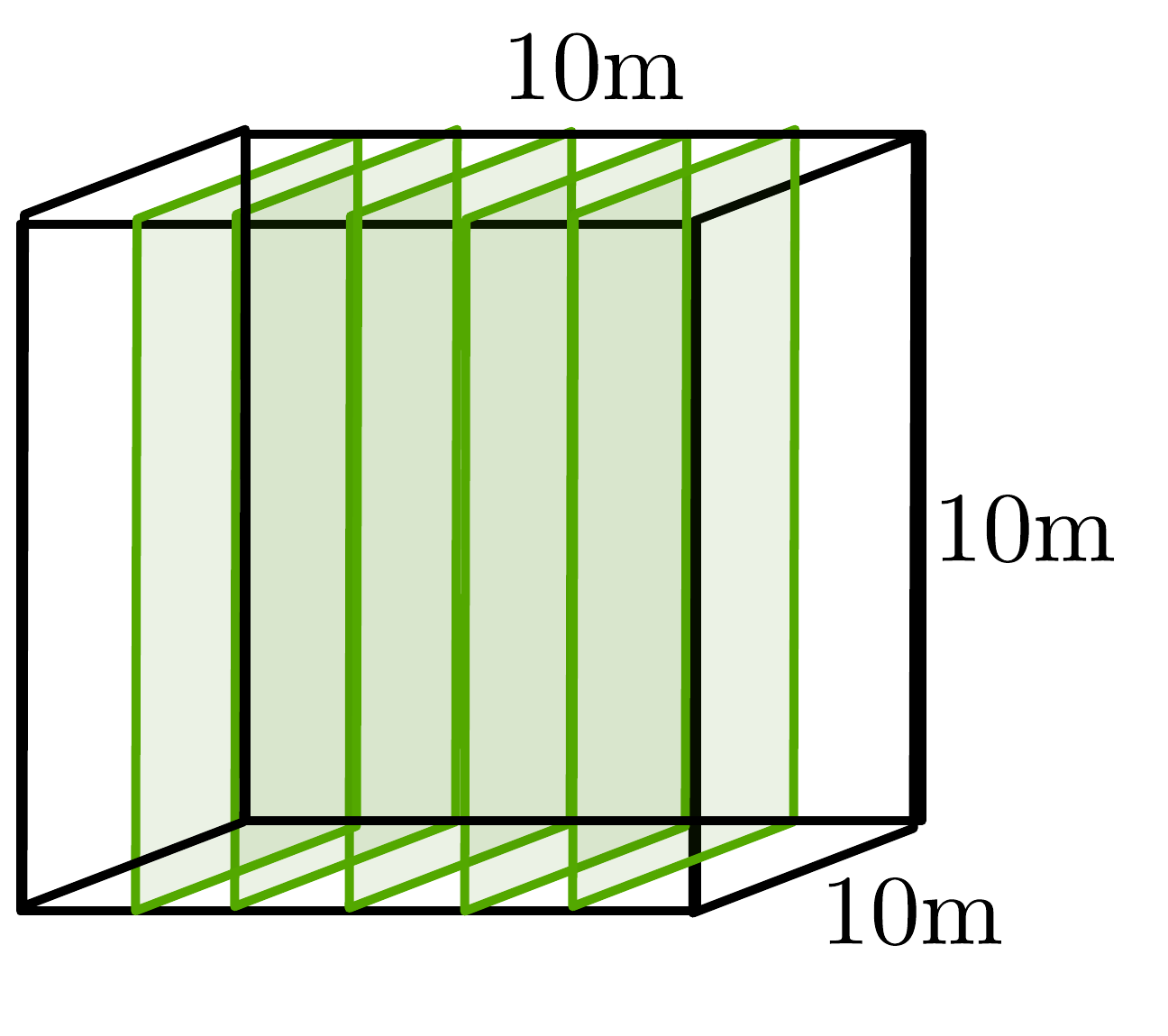} 
}
\subfigure[]{
\centering
    \includegraphics[width=0.15\textwidth]{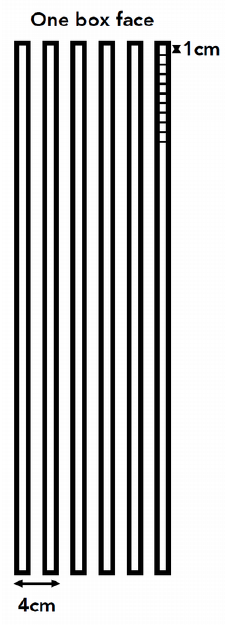}
}
\caption{\label{fig:nom_geo_cartoon}
    \cod nominal geometry: (a) box showing the $5\times$ inner stations and (b) the 6 layers for the face stations. 
}
\end{figure}

\subsection{Detector Description for High Energy Physics}
The geometry for both the \cod tracking layers as well as the two-scintillator configuration required for the background measurements in Sec.~\ref{sec:Measurement} is set up using the {\tt DD4hep}~\cite{dd4hep} toolkit. {\tt DD4hep} uses the {\tt ROOT} {\tt TGeometry} class for loading the detector geometry in memory and is being developed for the upgraded LHC detectors as well as future experiments. 

\subsection{Geometry construction for the full \cod detector}
\begin{figure}
\centering
\subfigure[]{
\centering
    \includegraphics[width=0.47\textwidth]{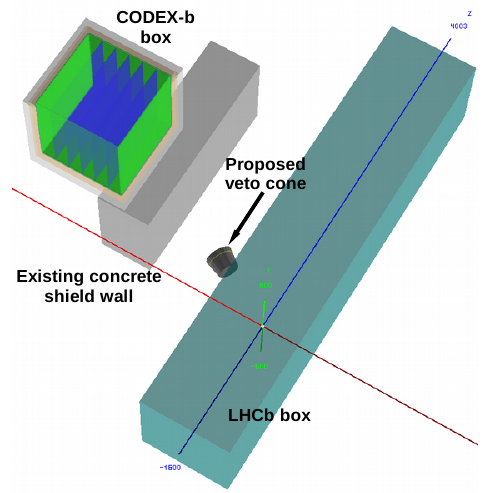} 
}
\subfigure[]{
\centering
    \includegraphics[width=0.47\textwidth]{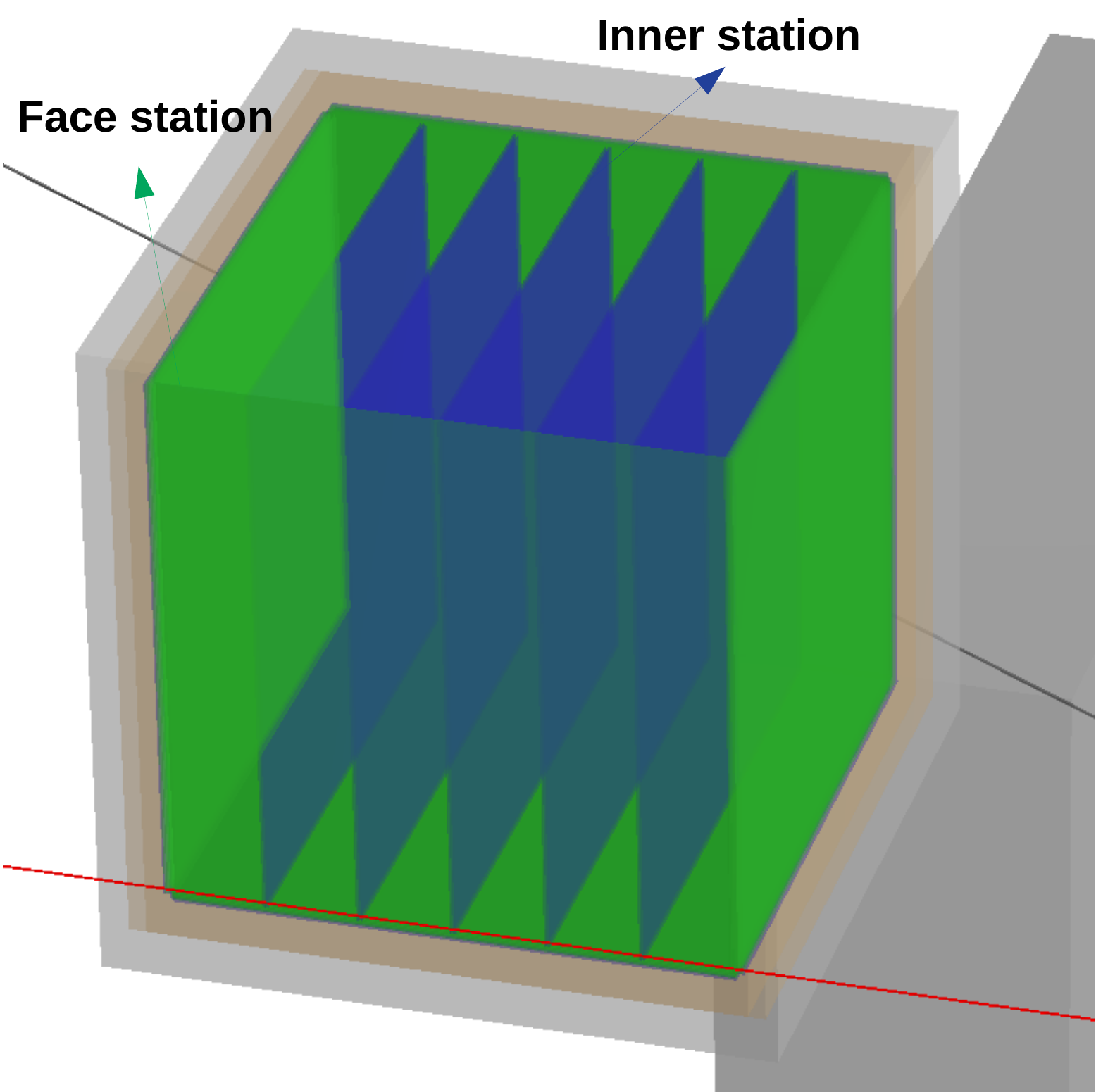}
}
\caption{\label{fig:geo_dd4hep}
    \cod simulation geometry in {\tt DD4hep}: (a) overall, (b) close-up view. 
}
\end{figure}

The envisioned detector geometry comprises two parts, face stations and inner stations. In the nominal version from the proposal~\cite{Gligorov:2017nwh}, there is a face station on each face of the \cod box volume, and each face station has 6 layers of resistive plate chambers (RPC) at 4~cm intervals with 1~cm granularity, as shown in Fig.~\ref{fig:nom_geo_cartoon}b. The size of each layer is $10 \times 10$~$m^{2}$ and the thickness is 2~cm. Further, the geometry includes 5 inner stations, as shown in Fig.~\ref{fig:nom_geo_cartoon}a, each containing a triplet of RPC layers. For simplicity, the RPC gas is replaced by Silicon tracker layers, with the hit timestamps saved from the simulation, since RPC's also provide timing information. The concrete shield wall with 3.2~m thickness, is placed just in front of \cod box. In addition, there is a proposed (not used in this analysis) veto cone~\cite{Gligorov:2017nwh} with two lead absorber and one active silicon layer sandwiched in between. Figure~\ref{fig:geo_dd4hep} shows the geometry construction in {\tt DD4hep}. 

The second geometry consists of two scintillator plates which is the same as our measurement configurations. The plastic material composition in {\tt GEANT} was adopted from HeRSCheL~\cite{LHCb-DP-2016-003} as polystyrene.

\begin{figure}
\centering
\includegraphics[width=\textwidth]{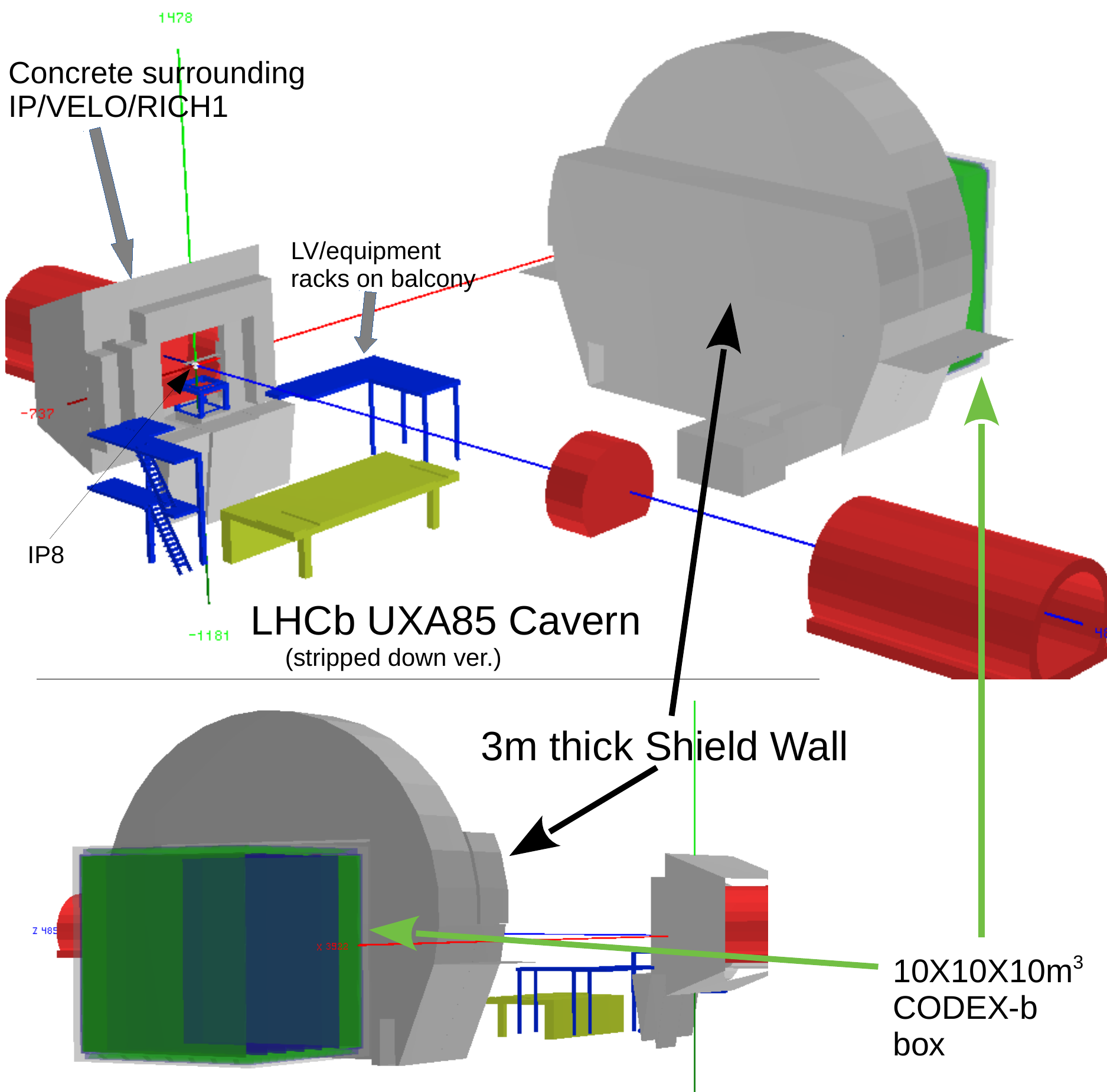} 
\caption{Cavern infrastructure incorporated in the simulation. The critical shielding elements include the concrete wall around the IP region and the shield wall. \label{fig:cavern_infra}}
\end{figure}

\begin{figure}
\centering
\includegraphics[width=0.7\textwidth]{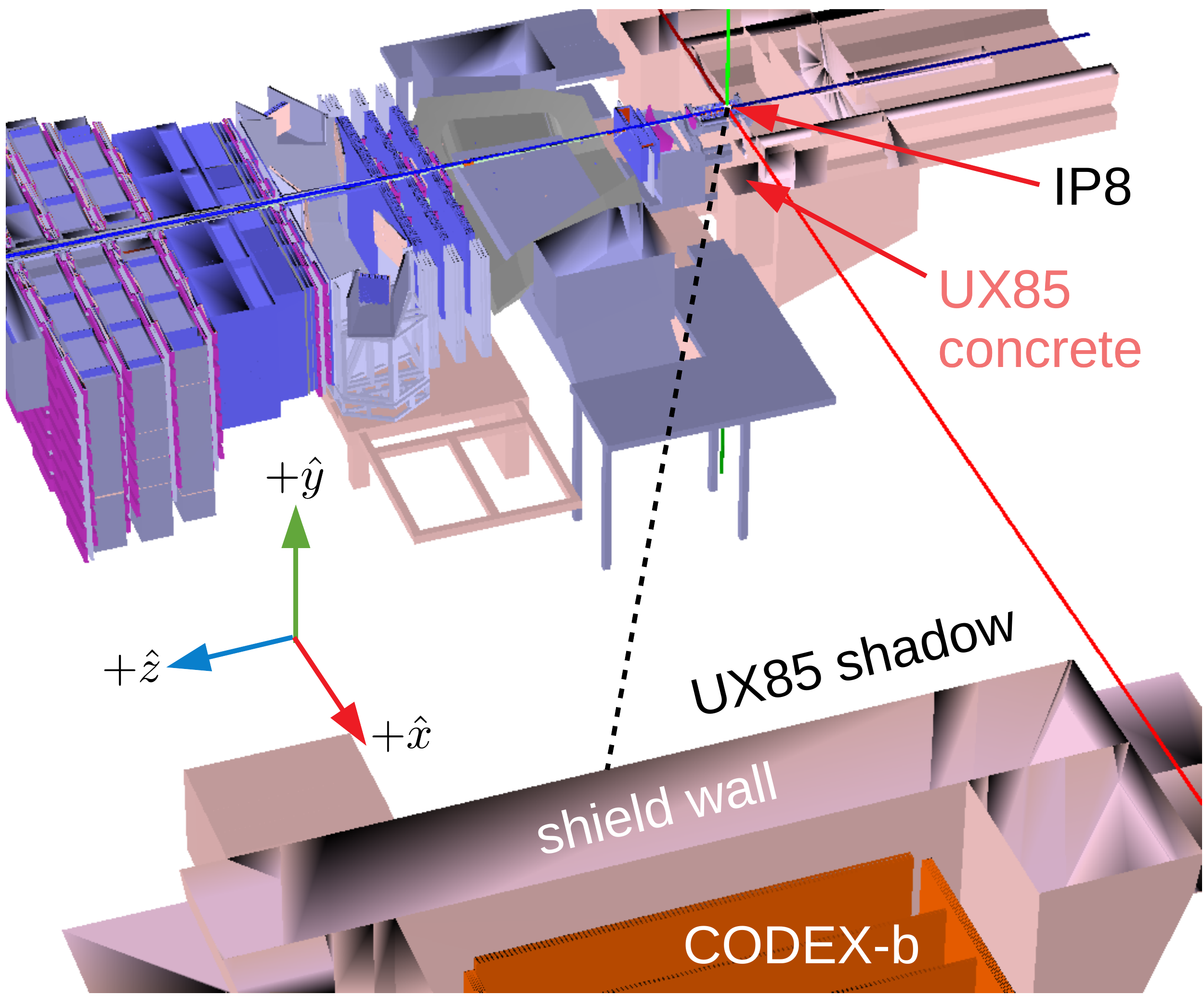} 
\caption{Horizontal cutaway/cross-section of the full geometry including the cavern infrastructure, LHCb detector, and \cod elements. The concrete protrusion marked as ``UX85 concrete'' leads to additional shielding in the \cod acceptance, but only for the more upstream region marked as ``UX85 shadow''. \label{fig:cavern_section_ux85shadow}}
\end{figure}

\begin{figure}
\centering
\includegraphics[width=0.65\textwidth]{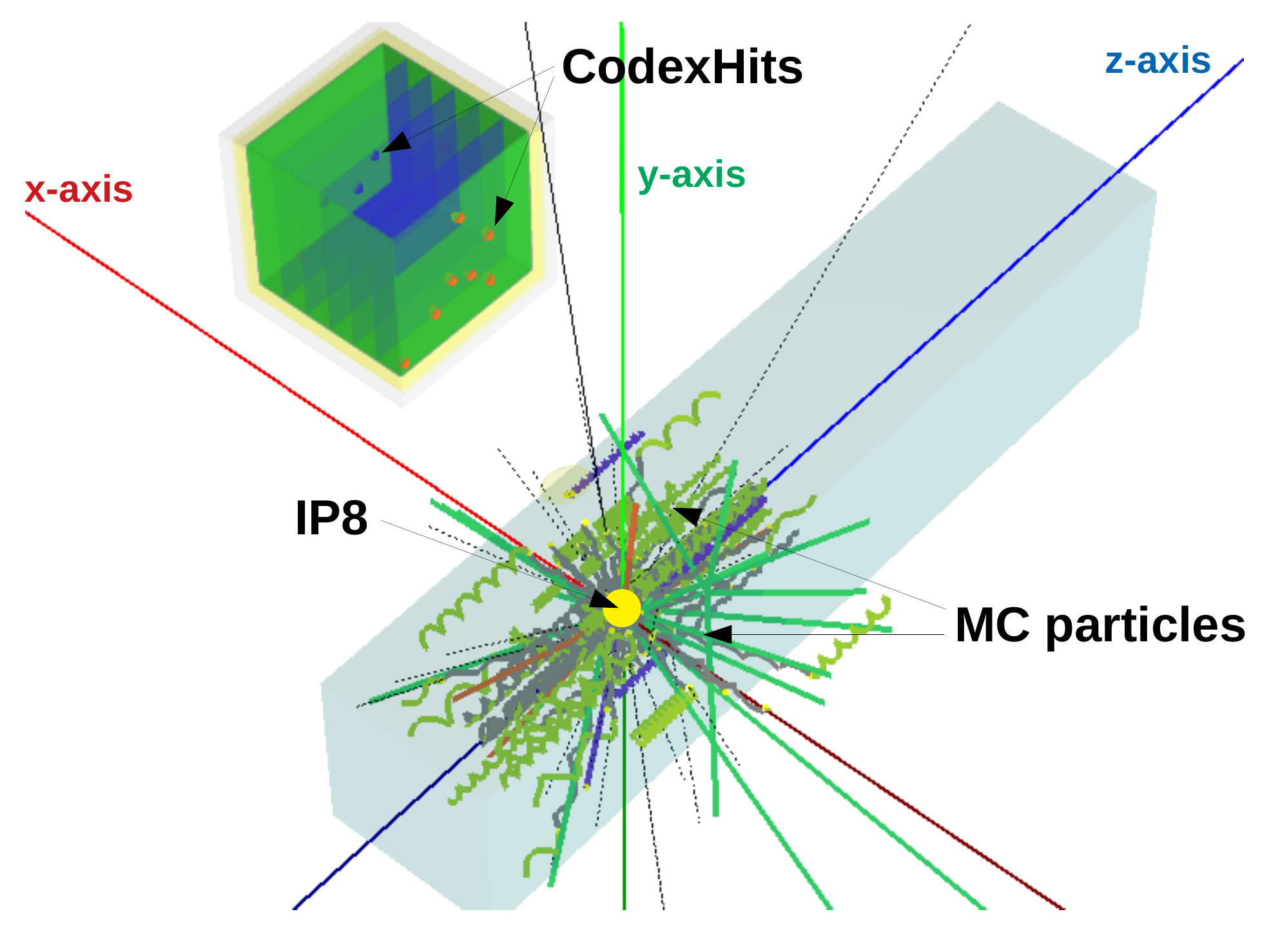}
\caption{\label{fig:dd4hep_events} A simple visual inspection of {\tt DD4hep} + {\tt DDG4} based simulation with minimum bias events, showing also the hits on the \cod plane.}
\end{figure}

\subsection{Cavern infrastructure}

Figure~\ref{fig:cavern_infra} shows elements of the cavern infrastructure included in the simulation. The cavern wall concrete geometry around the IP8 region has a rather non-trivial effect on the effective shielding geometry. Figure~\ref{fig:cavern_section_ux85shadow} shows a horizontal cutaway of the geometry, including the LHCb sub-detector elements, \cod, and the cavern infrastructure. The protrusion marked as ``UX85 concrete'' shields particles coming from the IP into the upstream region of the \cod volume. It does not have a wide enough coverage for the more downstream (higher $z$ or $\eta$) section of the \cod volume.

\subsection{Simulation setup in {\tt Gauss}}
\label{sec:Gausssetup}

The run conditions in the simulation were kept the same as in Run~II ($\sqrt{s}=13$~TeV proton-proton collisions) to match those of the background measurements. As a simple visual inspection, Fig.~\ref{fig:dd4hep_events} shows hits from minimum bias events with just the \cod volume in {\tt DD4Hep}. 

The final large simulation samples were generated using distributed computing resources made available through WLCG~\footnote{\href{https://wlcg.web.cern.ch/}{\tt https://wlcg.web.cern.ch/}}, interfaced via the {\tt Ganga} API~\footnote{\href{https://ganga.readthedocs.io/en/stable/index.html}{\tt https://ganga.readthedocs.io/en/stable/index.html}} within the LHCb software environment. The event generation and interfacing with {\tt Geant} was handled by the {\tt Gauss}~\cite{LHCb-PROC-2010-056,LHCb-PROC-2011-006}, the standard LHCb simulation package. The specific {\tt Gauss} version used corresponded to {\tt Geant4v10.4-patch-02} and {\tt Pythia8.2}~\cite{Sjostrand:2014zea} for generating the proton-proton collisions with minimum bias setting and specific LHCb tunings (for the beams, pileup, etc.). During Run~II, the beam-beam collision rate was 11.245 kHz per colliding bunch pair and there were 2332 such pairs, resulting in an overall collision rate of around 26~MHz. Each bunch crossing corresponded to $\approx1.1$ inelastic collisions. Therefore, approximately 29 million(M) generated events in the simulation corresponded to one second of recorded data, to match the Run~II data-taking conditions.

To save CPU time, the nominal tracking volume in LHCb simulations is restricted and does not extend till the shield wall. To include the \cod region in the DELPHI cavern, the tracking volume had to be extended to $[-5,40]$~m, $[-7,15]$~m and $[-5,25]$~m in the $x$, $y$ and $z$ coordinates, respectively (see Fig.~\ref{fig:cavern_section_ux85shadow} for the orientation of the axes). The  HeRSCheL scintillators as well as RPCs are MIP (minimum ionising particle) detectors, so that hits can be produced for very low energy tracks. Accordingly, the nominal LHCb simulation trackwise minimum energy cutoffs were lowered, especially to retain backgrounds from low energy neutron scattering. The low energy threshold for muon tracks were set to be $>1$~MeV, while, for all other species, they were set to $\sim 0$. The nominal {\tt Geant4} physics lists corresponded to the standard LHCb settings, including hadronic ({\tt FTFP\_BERT}), electromagnetic and general physics (synchrotron radiation, photo/electro-nuclear processes). Cherenkov processes for the RICH detectors (which were not required) were excluded, while {\tt DeltaRay} physics were included. The full LHCb detector, including the dipole magnet, the compensator and corrector magnets (see Fig.~\ref{fig:lhcb_magnets}) were included.

\clearpage
\subsection{Effects of the LHCb detector, magnetic fields and infrastructure elements}

\begin{figure}[h]
\centering
    \includegraphics[width=\textwidth]{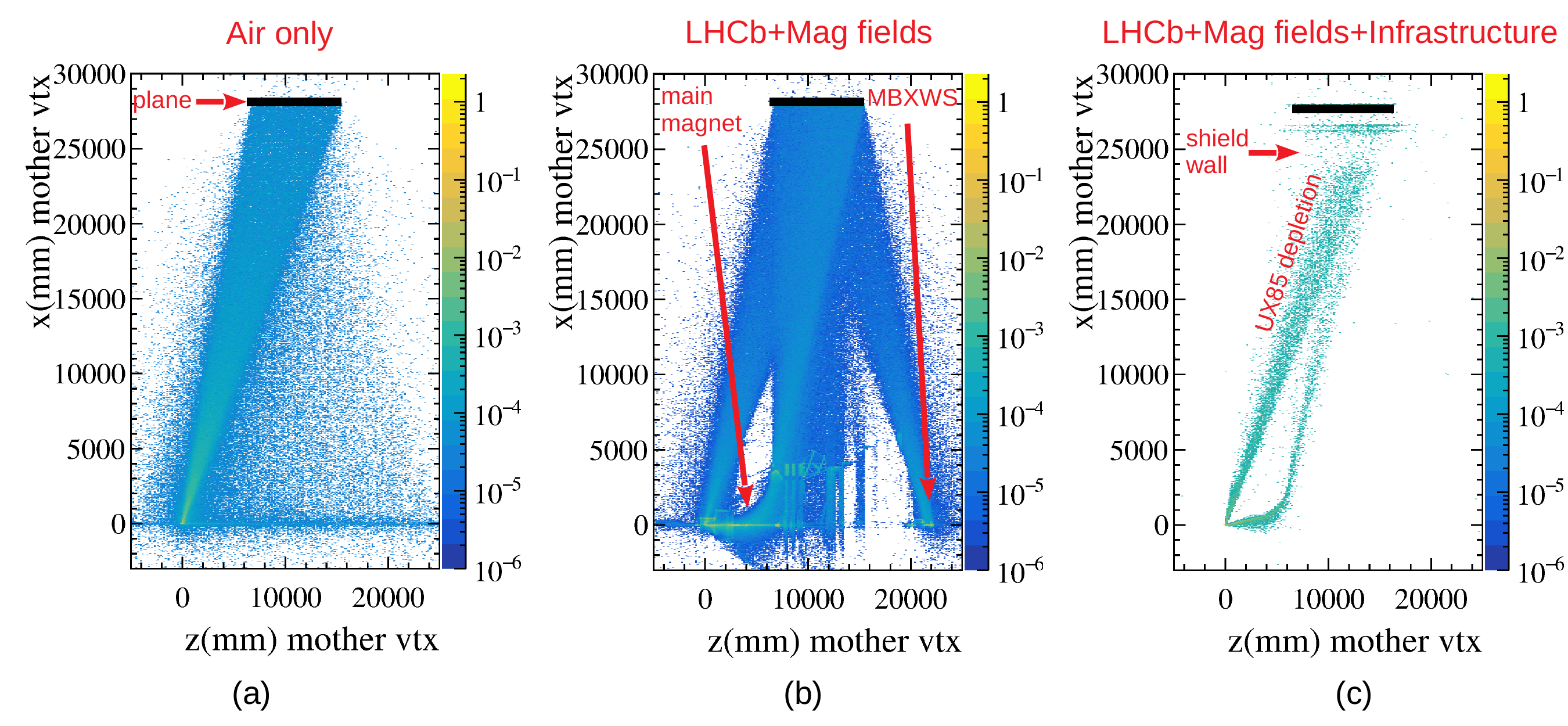}
\caption{\label{fig:twoPlane_OriVtx_distrib}
The origin vertices, derived from truth information in the simulation, of the tracks that result in hits on the plane corresponding to the measurement, behind the shield wall. The geometries correspond to (a) no structures, but only air (b) inclusion of the LHCb detector and various magnetic fields, (c) further inclusion of the cavern infrastructure elements, especially the shield wall.}
\end{figure}

Figure~\ref{fig:twoPlane_OriVtx_distrib} shows the distribution of the original vertices in the $x$-$z$ plane (looking from top, downwards), corresponding to the tracks that produce hits on the detector planes, kept in a similar position, as the background measurements at P1/P2/P4 in Fig.~\ref{fig:posconfig}. The origin vertex positions are derived from truth information in the simulation. The particles mostly correspond to $e/\mu/\pi/K$, with most of the muons from in-flight decays of $\pi/K$. The detector/infrastructure elements are incrementally introduced in this study, for illustration. Figure~\ref{fig:twoPlane_OriVtx_distrib}a shows the situation for proton-proton collisions in empty air. Figure~\ref{fig:twoPlane_OriVtx_distrib}b shows the effect of including the LHCb detector as well as the magnetic fields from the main LHCb and MBXWS corrector magnets behind the muon stations. The bending effect due to the magnetic fields are distinctly visible in Fig.~\ref{fig:twoPlane_OriVtx_distrib}b. The geometric position of the detector plane (marked as a thick black line in Figure~\ref{fig:twoPlane_OriVtx_distrib}) leads to a focusing effect of the magnetic fields, in momentum. That is, the LHCb magnet bends tracks with a momentum of $\approx 1$~GeV, into the DELPHI region. Similarly, the momentum of the tracks bent by the MBXWS corrector magnets into the DELPHI side is $\approx 600$~MeV.

\begin{figure}[h]
\centering
\subfigure[]{
\centering
    \includegraphics[width=0.45\textwidth]{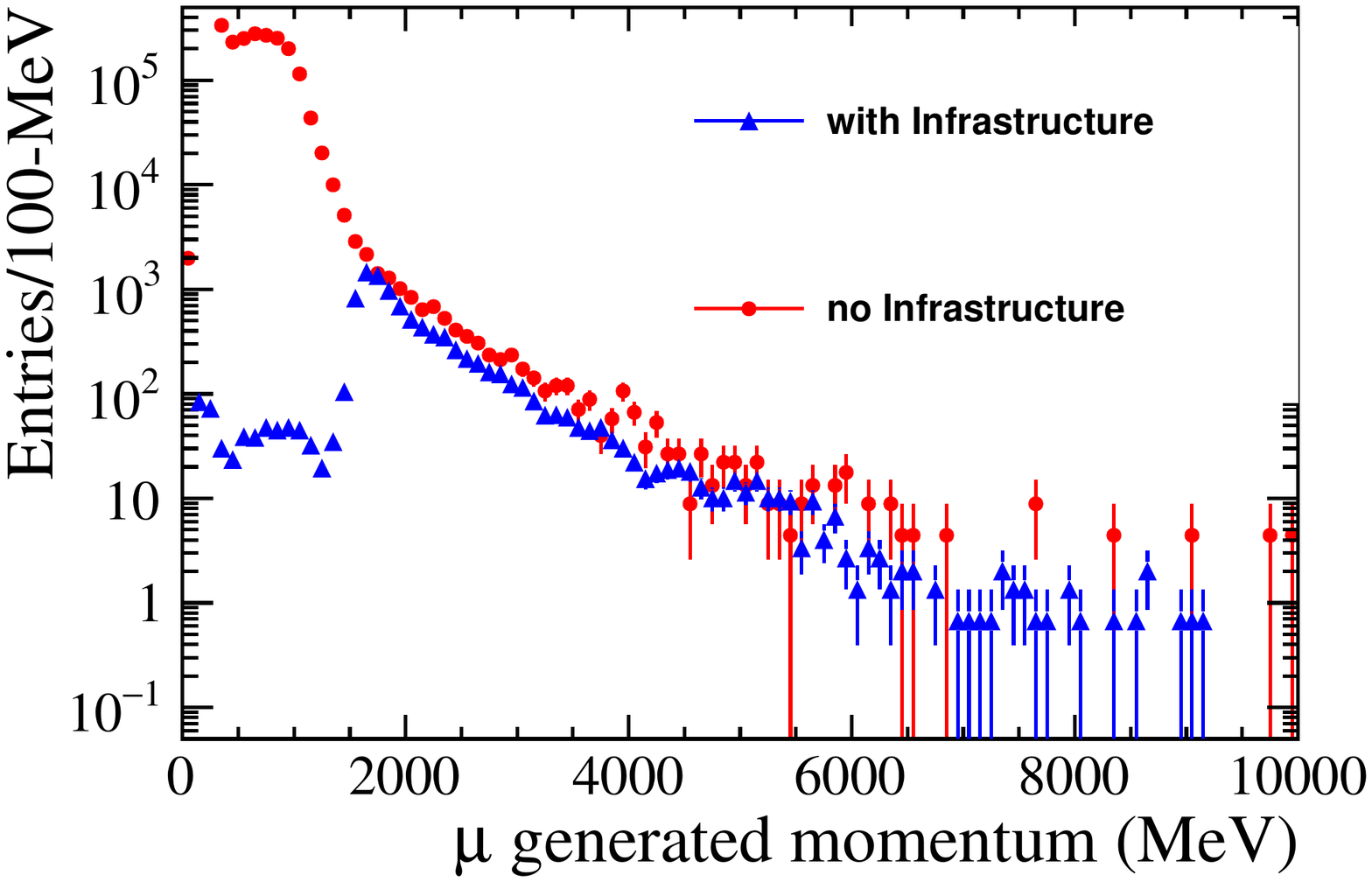}
}
\subfigure[]{
\centering
    \includegraphics[width=0.45\textwidth]{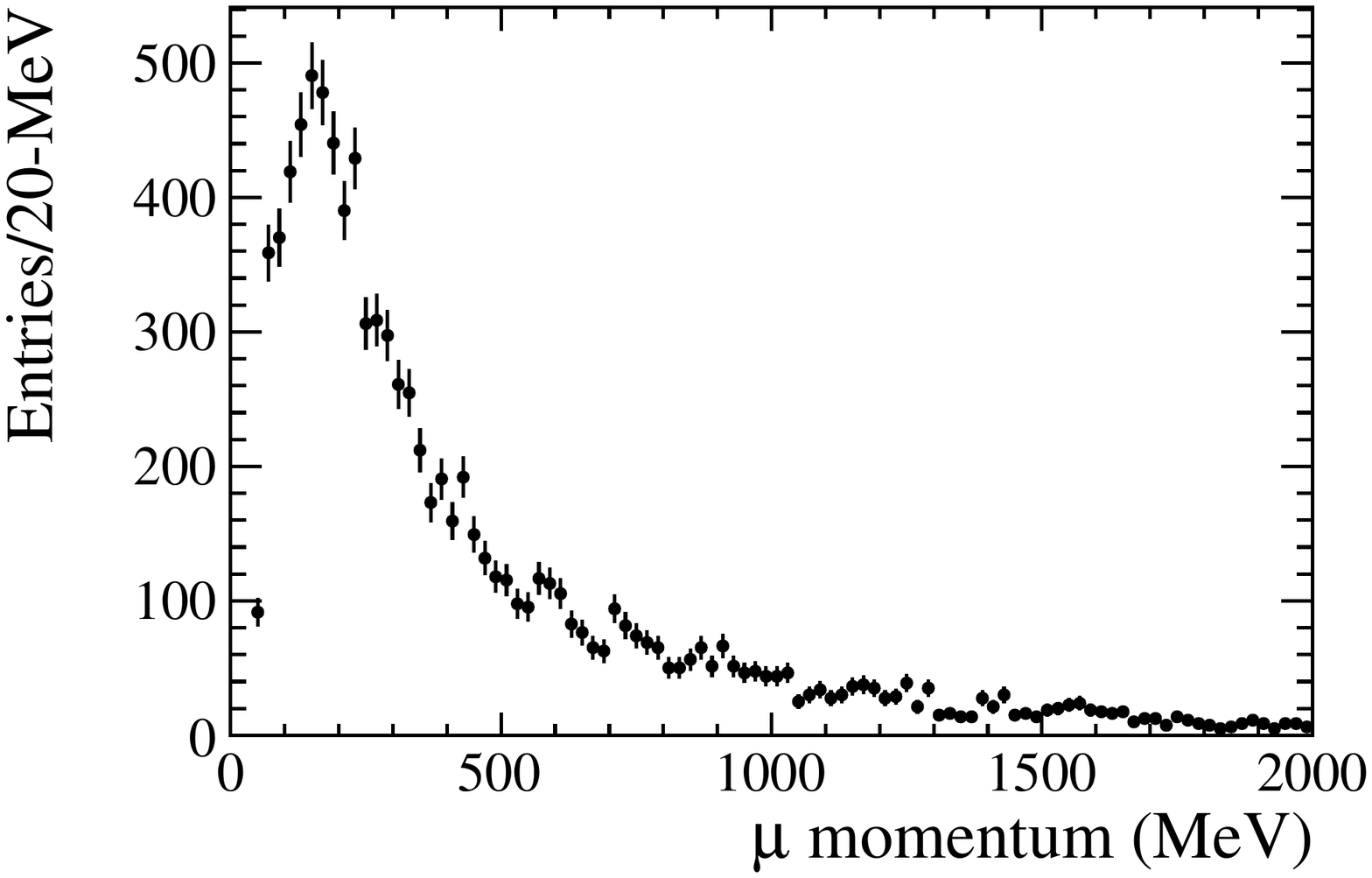}
}
\caption{\label{fig:gen_mom_distrib_muons}
The momentum distribution of muons producing hits on a detector plane behind the shield wall: (a) initial momentum at production without and with the cavern infrastructure elements; muons generated with momentum less than $\sim 1500$~MeV are stopped by the shields. (b) Final momentum after all energy losses showing that the slowed-down muons have quite low momenta.}
\end{figure}

Figure~\ref{fig:twoPlane_OriVtx_distrib}c shows the distributions after including the cavern infrastructures with the various concrete shielding elements. The depletion due to the ``UX85 shadow'' in Fig.~\ref{fig:cavern_section_ux85shadow} is visible. The slower bent tracks from the MBXWS magnetics are completely shielded off. To understand, which tracks are stopped and which make it through the shield wall, Fig.~\ref{fig:gen_mom_distrib_muons}a shows the distribution of the muon momentum at generation of the tracks that ultimately produce hits on the detector planes. In red is the distribution without the infrastructure (corresponding to Fig.~\ref{fig:twoPlane_OriVtx_distrib}b), while the blue triangles correspond to Fig.~\ref{fig:twoPlane_OriVtx_distrib}c, including the full cavern infrastructure. From Fig.~\ref{fig:gen_mom_distrib_muons}a, the concrete wall stops tracks with momentum below $\approx 1500$~MeV, consistent with expectations from a 3.2~m thick concrete barrier. Figure~\ref{fig:gen_mom_distrib_muons}b shows the distribution of slowed-down muon momenta at the detector plane position, after energy losses. Aside a long tail at high momenta, the distribution peaks around $\sim 200$~MeV.

\subsection{Final hit rate as a function of $z$}
\label{sec:hitrate_x}

\begin{figure}[h]
\centering
    \includegraphics[width=0.66\textwidth]{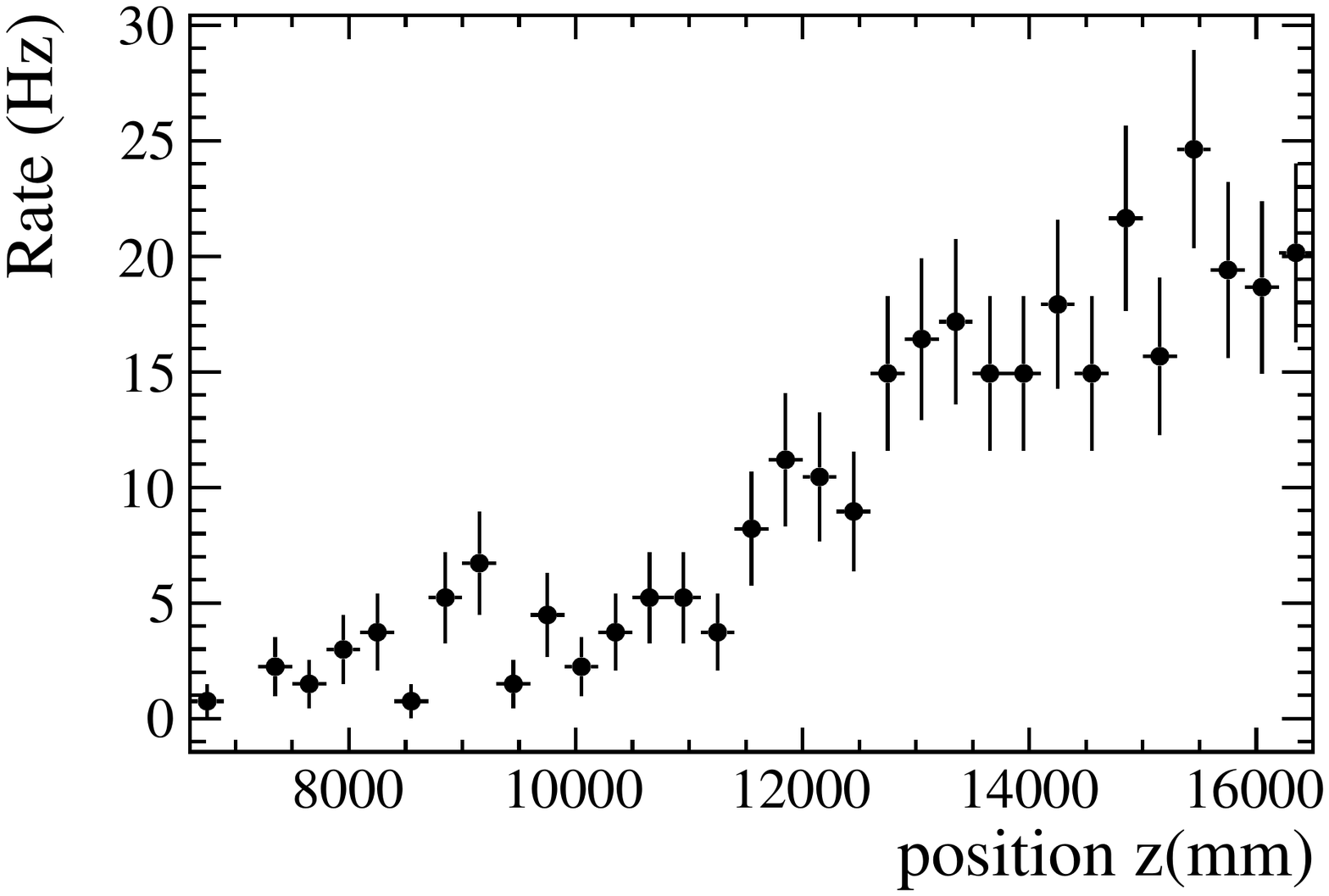}
\caption{\label{fig:final_result_notE_hitrate}
The hit rate from the simulation across a vertical plane parallel to the beam line and area $30\times30$~cm$^2$, at $y=0$ and varying $z$ position, corresponding to the measurement configurations P1$\to$P2$\to$P4 in Table~\ref{table:rate_stable_beam}.} 
\end{figure}

Figure~\ref{fig:final_result_notE_hitrate} shows the  hit rate from the simulation across a vertical plane parallel to the beam line, normalized to the scintillator area $30\times30$~cm$^2$. The rates are at $y=0$ (at the height of the beamline) and varying $z$ (from upstream to downstream) position, corresponding to the measurement configurations P1$\to$P2$\to$P4 in Table~\ref{table:rate_stable_beam}. The simulation sample consists of around 19 million minimum bias events generated with the full geometry and includes the trigger aspects for the data from Sec.~\ref{sec:trigger}.

The trend that the rate increases with increasing $\eta$ (from upstream $z$ to downstream $z$), as observed in Table~\ref{table:rate_stable_beam} for the data is shown by the simulation. However the simulation overestimates the rate compared to the data by over an order of magnitude. At the largest $z$ position, the measured hit rate over a $30\times30$~cm$^2$ area was around 0.5~Hz, while the simulation gives a rate of around 20~Hz. 

The distributions from the simulation in Figs.~\ref{fig:gen_mom_distrib_muons} and \ref{fig:final_result_notE_hitrate} using the {\tt Pythia8}~\cite{Sjostrand:2014zea} generator were validated against an independent FLUKA~\cite{Ferrari:898301} study using the DPMJET-III generator~\cite{10.1007/978-3-642-18211-2_166} (private communication, F.~Cerutti and A.~Ciccotelli, CERN EN-STI group). 

Inside LHCb's forward acceptance, the occupancies in the Outer Tracker sub-detector were also also found to be of the same order as predicted by the {\tt Pythia8} simulation.

\subsection{Additional shielding elements in \cod acceptance not included in the simulation}
\label{sec:additional_mat}

\begin{figure}
\centering
\includegraphics[width=1\textwidth]{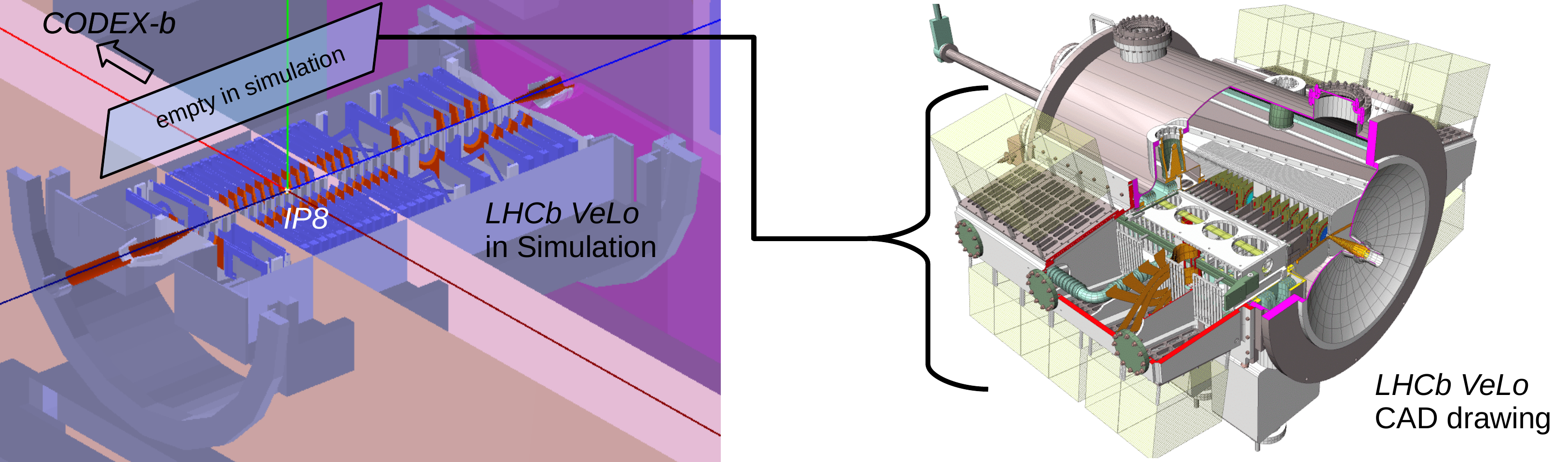}
\caption{\label{fig:velo}
   Additional material on the sides of the VeLo not included in the simulation (left), visible in the representative CAD drawing (right). The materials on the left side are entirely in \cod's acceptance.}
\end{figure}

A precise description of the cavern infrastructure pertinent to \cod's acceptance is difficult to obtain. The \cod acceptance at high transversity is completely disjoint with LHCb's forward acceptance, and therefore the effective radiation length of the material within this particular acceptance is poorly known. 
One additional element not included in the simulation is material on the sides of the current (for Run~I and Run~II data-taking) VeLo as shown in Fig.~\ref{fig:velo}. This includes~\footnote{\href{https://www.nikhef.nl/pub/departments/mt/projects/lhcb-vertex/index\_current.html}{\tt https://www.nikhef.nl/pub/departments/mt/projects/lhcb-vertex/index\_current.html}} the hood and flanges for the secondary vacuum vessel, cooling and electronics cables, and the readout boards. The precise amount of radiation length offered by these materials is difficult to infer from the drawings, and will also change, moving to the upgraded VeLo. The VeLo side wall composed of stainless steel is estimated to be at least 3~cm thick (private communication, F.~Sanders, LHCb VeLo group). 


\begin{figure}
\centering
    \includegraphics[width=0.5\textwidth]{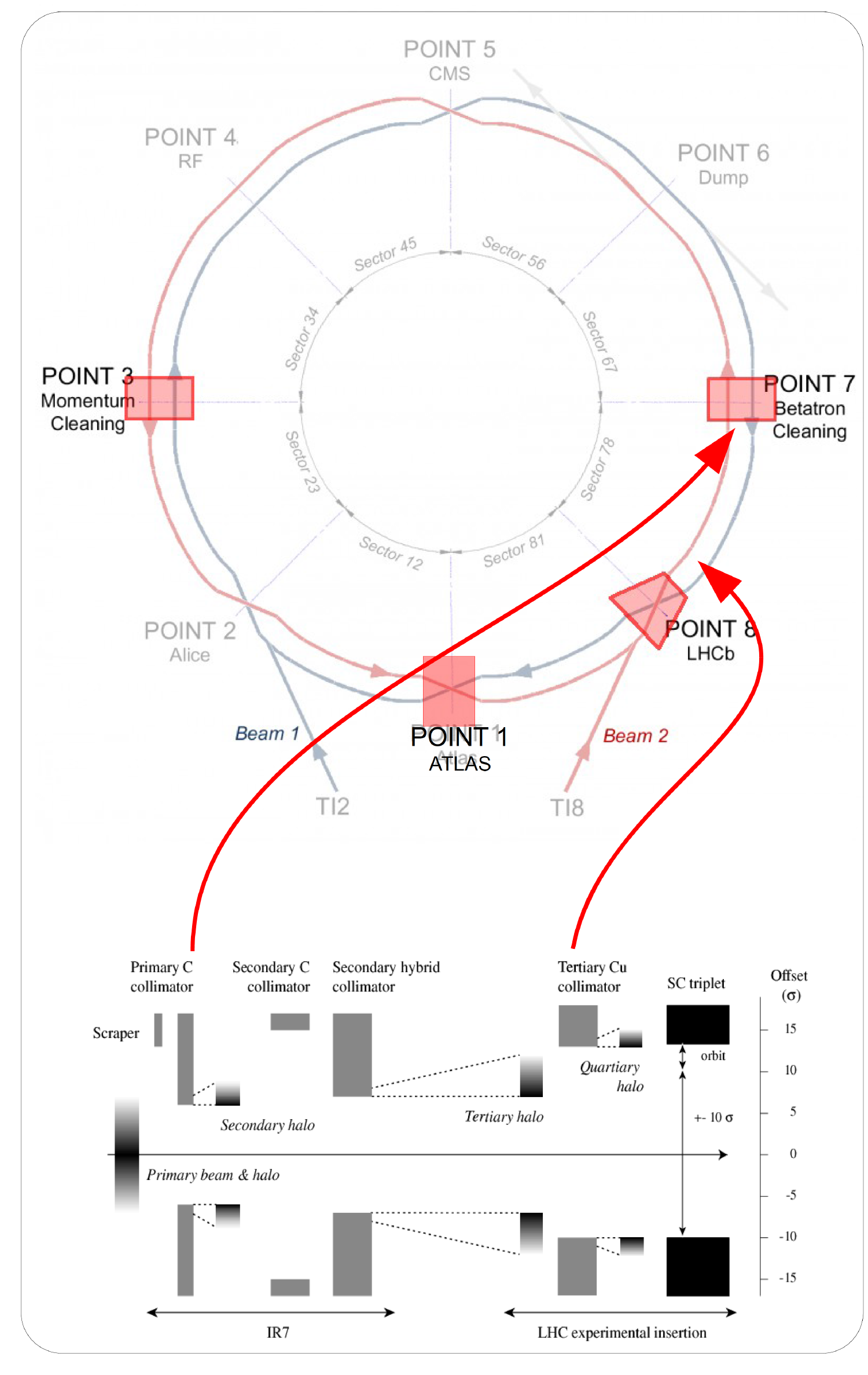} 
\caption{\label{fig:lhc_ring} The LHC, and various interacting regions (IRs).}
\end{figure}

\begin{figure}
\centering
    \includegraphics[width=0.9\textwidth]{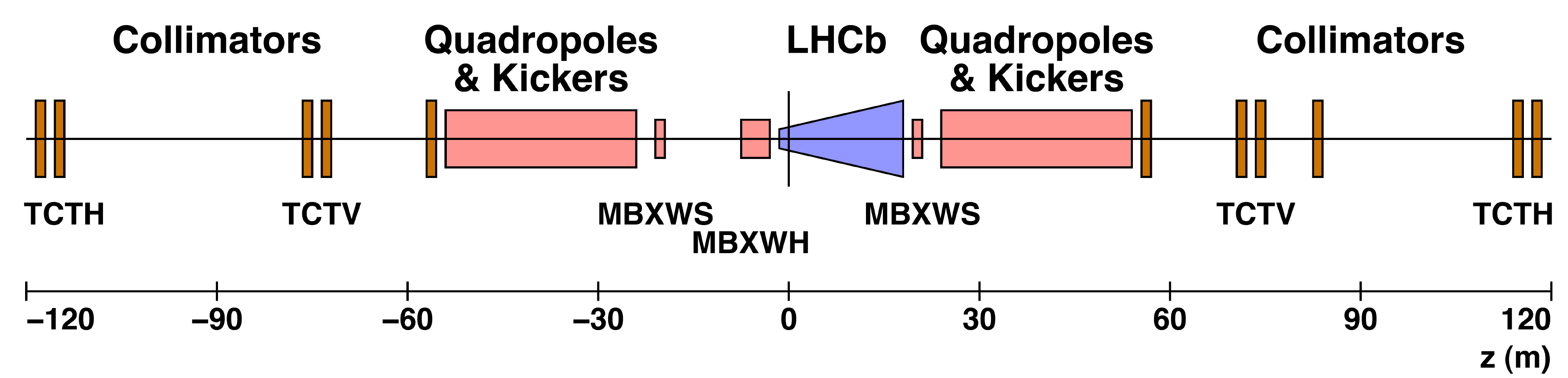} 
\caption{\label{fig:lhcb_magnets} The various magnets around the LHCb region relevant to this study.}
\end{figure}

\subsection{The LHC machine induced background}
\label{sec:mib}

For background studies, the flux source is not only the proton-proton collisions at IP8, but also proton losses from the LHC machine itself. The rate of this machine induced background (MIB) is generally proportional to the machine beam current and depends on the machine optics and vacuum conditions, among others. A detailed description of the MIB sources and simulation tool is described in Refs.~\cite{Appleby:1376692,Lieng_thesis}. The sources relevant to the LHCb cavern include global sources resulting in losses on the IR8 Tertiary Collimators (TCTs) (see Fig.~\ref{fig:lhc_ring} and Fig.~\ref{fig:lhcb_magnets}):
\begin{itemize}
\item IR7 Betatron cleaning of the transverse beam halo components
\item IR3 momentum cleaning (taken to be small for LHCb)
\item Elastic beam-gas interactions (BGI)
\item IP1 ATLAS (much higher luminosity) elastic interactions leading to showers at IP8,
\end{itemize}
as well as local sources that give direct showers inside the cavern:
\begin{itemize}
\item BGI along the Long Straight Section leading up to the LHCb experiment (LSS8)
\item BGI in LHC arcs with losses at the IR8 TCTs near LHCb.
\end{itemize}
The beam-gas losses originate from proton beam interactions with the gas residue in the vacuum chamber of the LHCb region. For LHCb, Beam~1 is more important than Beam~2 due to trigger (on-time vs. off time) and this is likely to remain true for \cod, keeping in mind that the latency from IP8 is of the same order as for the last muon station in LHCb.

\begin{figure}
\centering
    \includegraphics[width=0.7\textwidth]{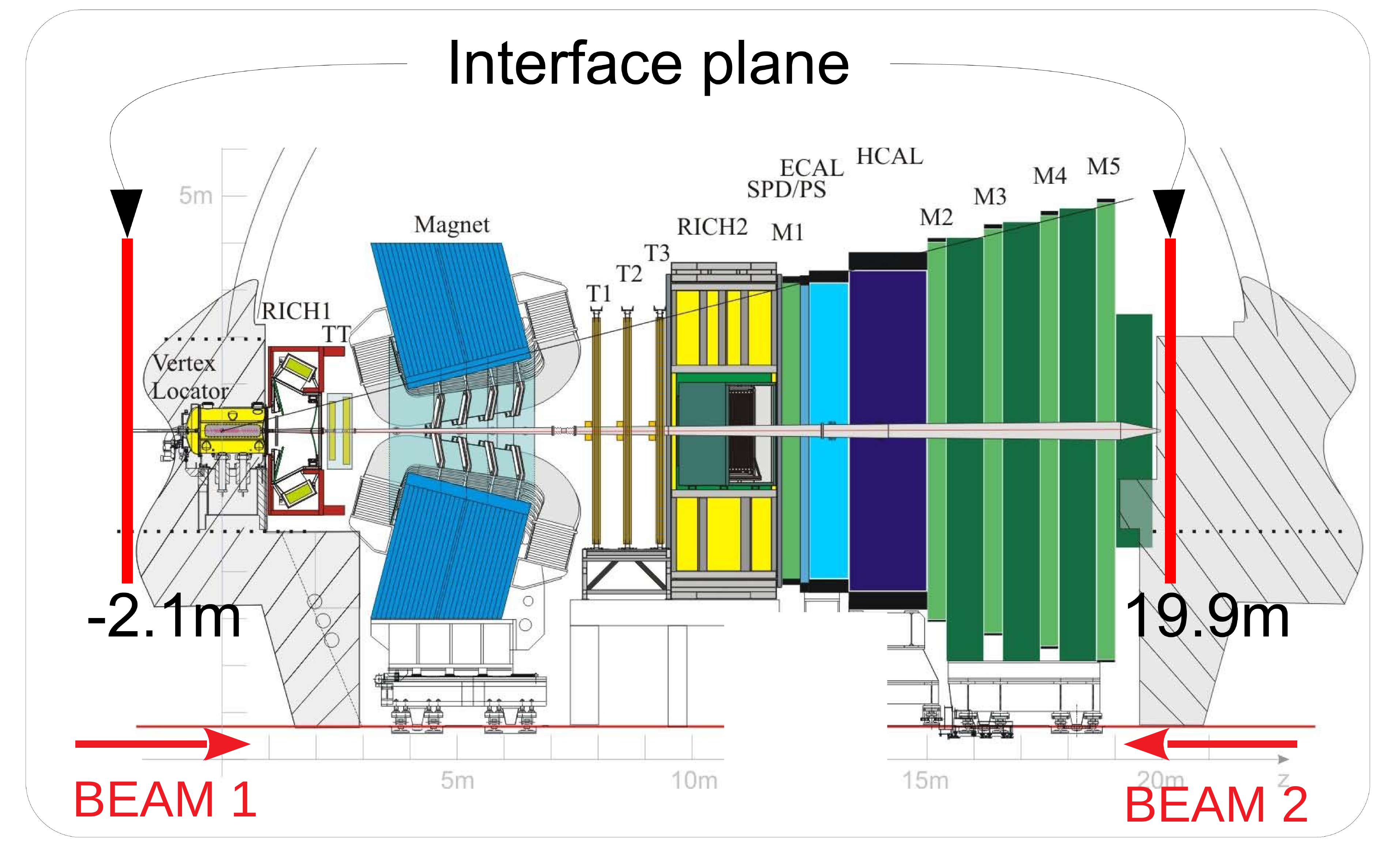} 
\caption{\label{fig:interfaceplanes} The two effective interface planes for MIB sources for Beam~1 and Beam~2 with respect to the current LHCb detector.}
\end{figure}

\begin{figure}
\centering
\subfigure[]{
\centering
    \includegraphics[width=0.4\textwidth]{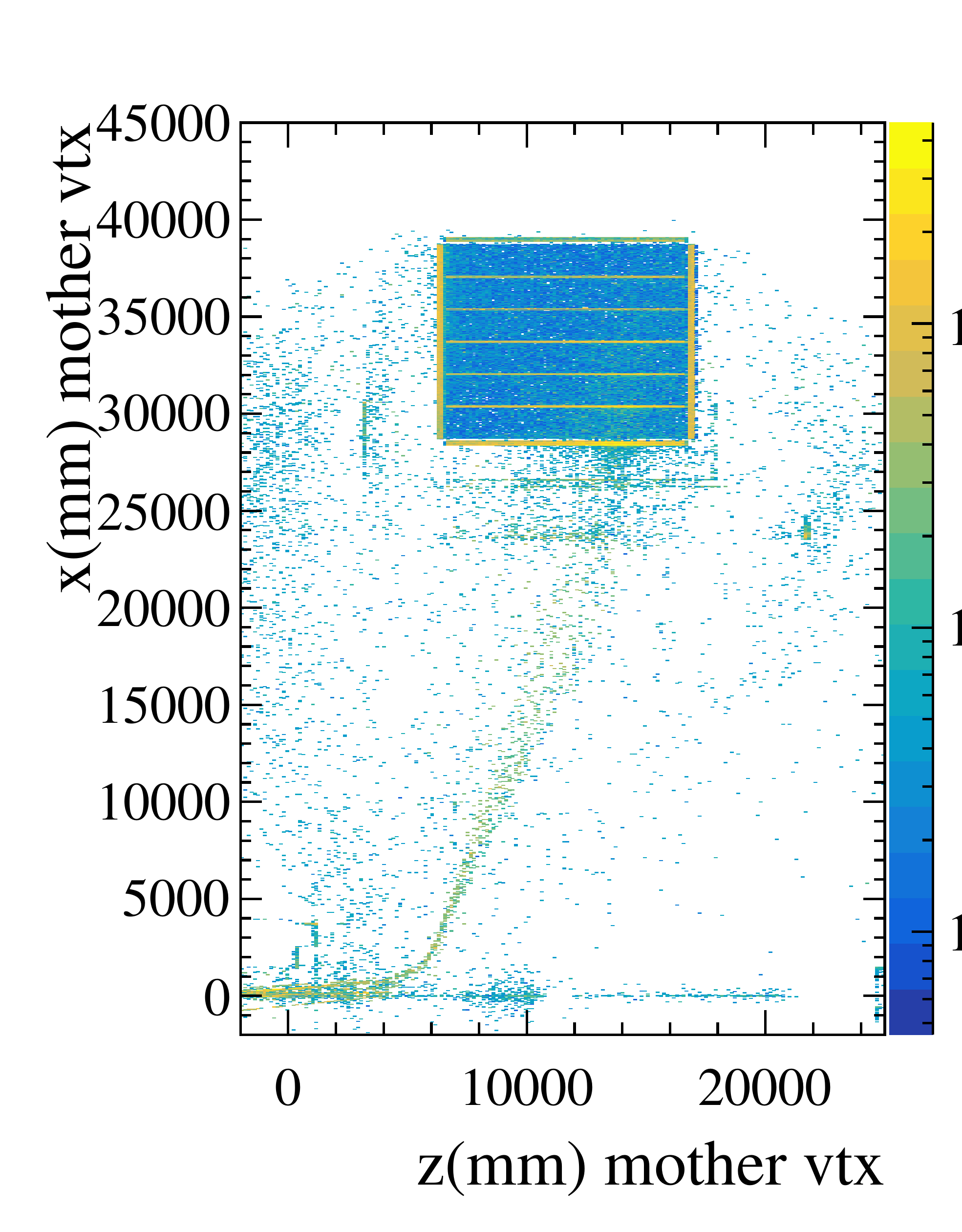} 
}
\centering
\subfigure[]{
\centering
    \includegraphics[width=0.4\textwidth]{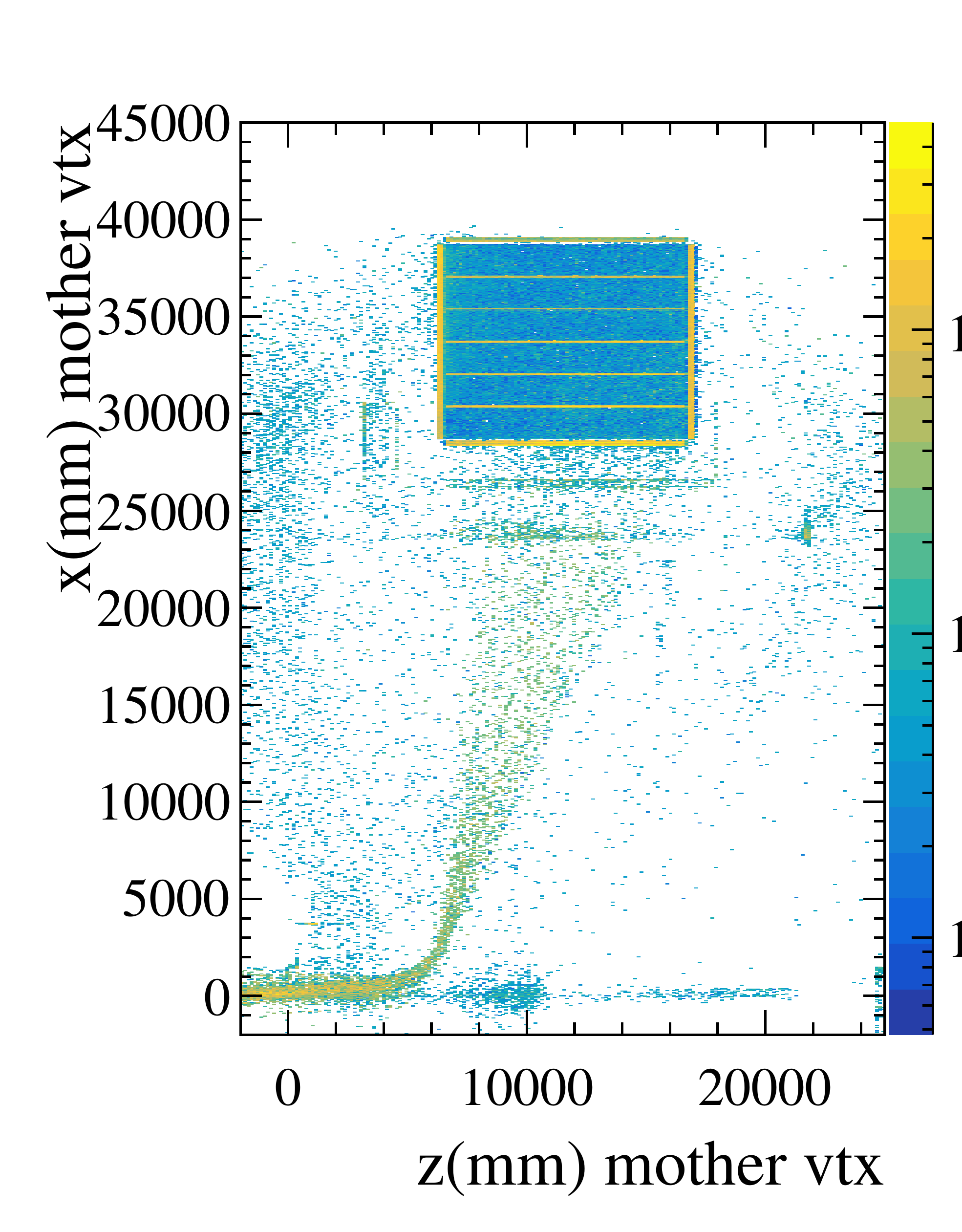} 
}
\centering
\subfigure[]{
\centering
    \includegraphics[width=0.4\textwidth]{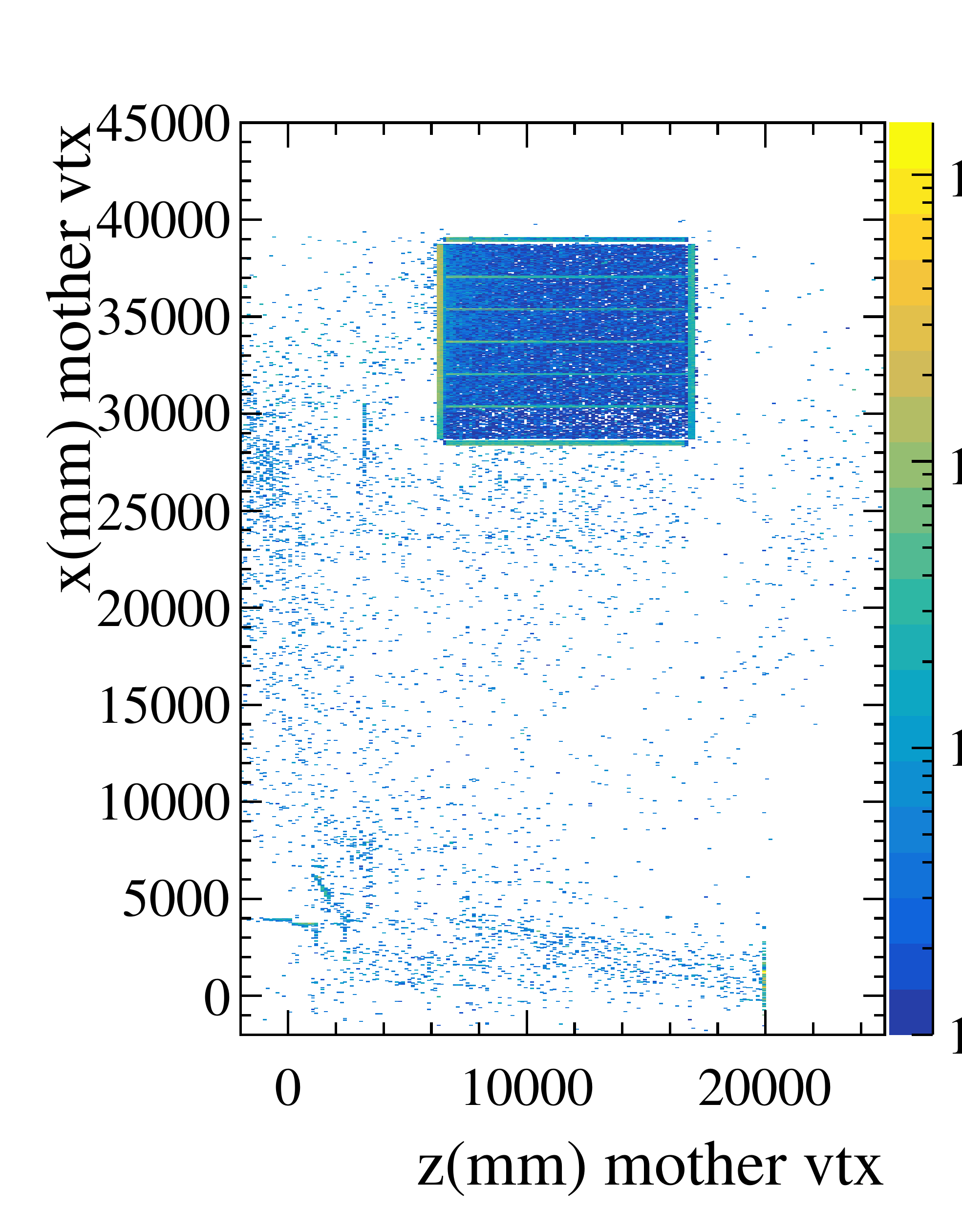} 
}
\centering
\subfigure[]{
\centering
    \includegraphics[width=0.4\textwidth]{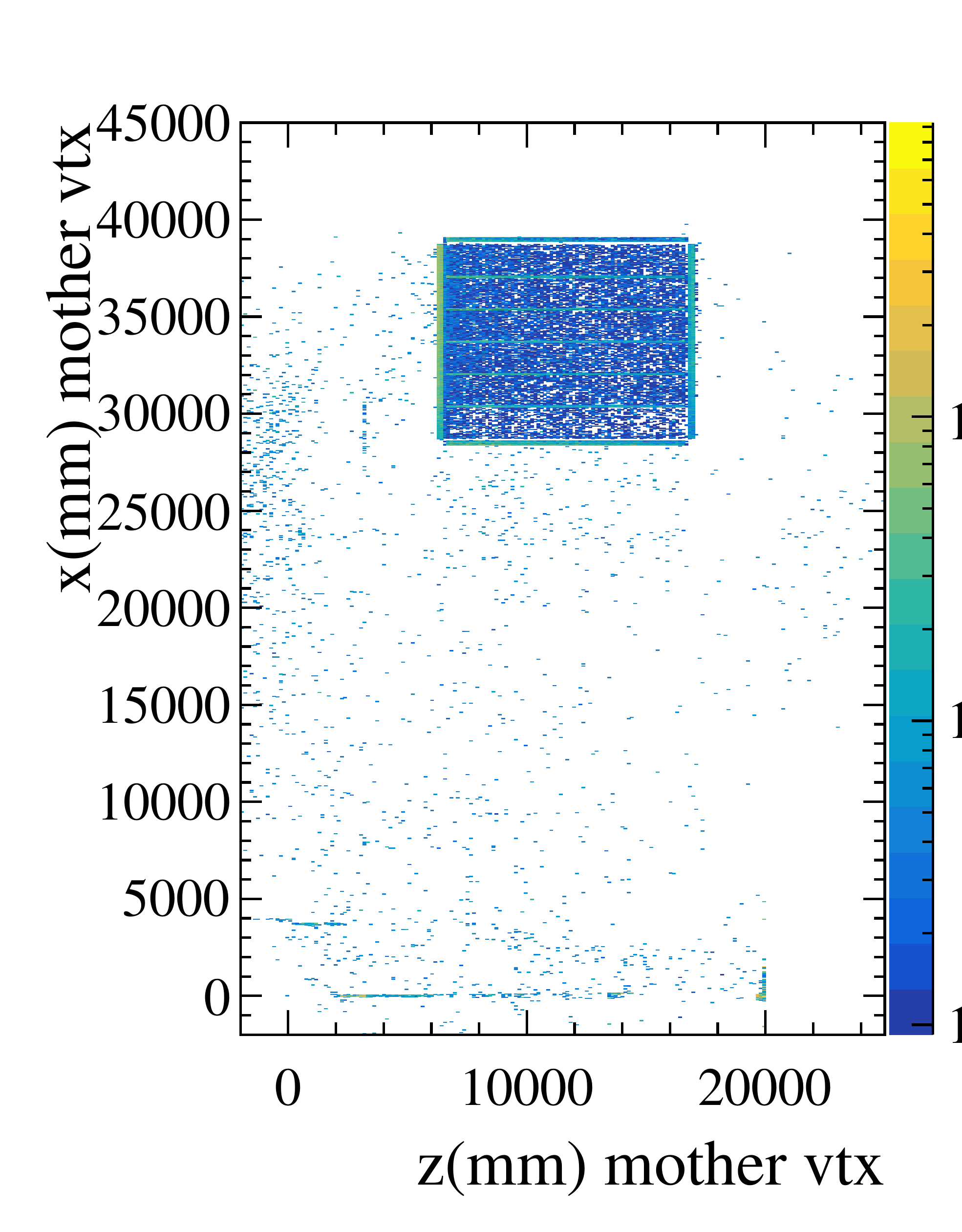} 
}
\caption{\label{fig:oriVtx_MIB}
    $x$-$z$ distribution of the decay vertex generating the particles that produce hits on the \cod sensitive detector: (a) Beam1 BetatronCleaningTCT (b) Beam1 BeamGasLSS (c) Beam2 BeamGasArcTCT and (d) Beam2 BetatronCleaningTCT. Note the effect of the LHCb dipole magnetic field for the Beam~1 sources (a) and (b).
}
\end{figure}

\begin{table}
\begin{center}
\begin{tabular}{c|c } 
 \hline
 Description & $N_{\rm gen}$ \\ \hline \hline
 Beam~1, IR7 Betatron cleaning TCTV &  418~M\\ 
 Beam~1, BGI losses along LSS8 & 19.4~M \\ 
 Beam~2, elastic BGI in LHC arcs with TCTV losses & 2440~M \\ 
 Beam~2, IR7 Betatron cleaning inefficiencies at TCTV & 96.5~M \\ 
 \hline
\end{tabular}
\end{center}
\caption{Generator-level simulation statistics for the MIB background studies. \label{tab:stats}}
\end{table}

The simulation tool transports the resulting showers from the various sources to two interface planes that act as effective sources as shown in Fig.~\ref{fig:interfaceplanes}. The simulation samples deemed most relevant for the LHCb cavern are listed in Table~\ref{tab:stats}, using the full \cod volumes instead of the two-plane background measurement configuration. Figure~\ref{fig:oriVtx_MIB} shows the $x$-$z$ distribution of the origin vertices of the particles that leads to hits. For the two Beam~1 sources, the showers feel the LHCb magnetic field, while the Beam~2 sources, the direction is not favorable. From the discussion in Sec.~\ref{sec:additional_mat}, even without the MIB component, the simulation overestimates the background flux. Once this discrepancy is understood, the additional MIB contribution due to Beam~1 sources will be investigated.

%% file: chapters/summary.tex
\section{Summary and ongoing work}
\label{sec:Summary}

In summary, a successful background measurement campaign was held in August 2018 to measure the radiation flux in the \cod volume area during LHC data taking conditions. The average hit rate under stable beam condition was found to be much higher than the average hit rates of pure ambient background. Therefore, pure ambient background when the beam is off, can be neglected. The maximal background rate when the beam is on, just behind the concrete shield wall, was around 0.6~mHz/cm$^{2}$. The measurements also demonstrated a strong $\eta$ dependence of the background rate.

A simulation suite incorporating the full LHCb detector, magnetic fields and cavern infrastructure geometry, within the standard LHCb simulation software, {\tt Gauss}, is reported. The simulation reproduces the $\eta$ dependence seen in the measurement data, but overestimates the normalization. Possible shielding elements missing in the simulation are pointed out. While the setup presented here describes the Run~1-2 LHCb detector, it can be trivially extended to the Upgrade configuration. Preliminary studies of non-collision LHC machine induced backgrounds are also reported.

The simulation demonstrates that even with known shielding elements excluded from the geometry, the background flux rate is overestimated, compared to the measurements. The study therefore validates that the background estimates for the physics reach projections in Ref.~\cite{Gligorov:2017nwh,Aielli:2019ivi} are conservative. Further development of the simulation suite remains ongoing on several fronts. The missing shielding elements are to be better understood. The generation and propagation of rare hard processes not currently included within the current framework have to be facilitated. Tracking studies with the full \cod detector and signal decays, as reported in Ref.~\cite{Aielli:2019ivi}, are also ongoing.

%% file: acknowledgements.tex
\section*{Acknowledgements}

\noindent We acknowledge the support and help from Federico Alessio, Gloria Corti, Raphael Dumps, Markus Frank, Vava Gligorov, Simon Knapen, Tengiz Kvaratskheliya, Dominik M\"{u}ller, Dean Robinson, Heinrich Schindler and Niels Tuning at various stages of this study. We thank the overall computing and simulation teams of the LHCb collaboration for their generous help in performing the simulation studies presented here. We thank the LHCb technical coordination for making the background studies in the DELPHI cavern possible. We thank the LHCb management, especially Giovanni Passaleva and Chris Parkes, for advice, support and encouragement on the project. J. L. and C. S. M. is supported by the National Research Foundation of Korea (NRF) grant funded by the Korea government (MSIT) (Grants No. 2018R1A6A1A06024970 and No. 2018R1C1B5045624).